\documentclass{jfm}
\usepackage{graphicx}
\usepackage{float}
\usepackage{epstopdf, epsfig}
\usepackage{caption,subcaption}
\usepackage[intlimits]{amsmath}
\usepackage{lscape}

\newcommand\Web{\mbox{\textit{We}}}
\newcommand\Mar{\mbox{\textit{Ma}}}
\newcommand\Fro{\mbox{\textit{Fr}}}
\shorttitle{Falling films with surfactants}
\shortauthor{A. Batchvarov et al.}

\title{Three-dimensional dynamics of falling films in the presence of insoluble surfactants}

\author{Assen Batchvarov\aff{1},
  Lyes Kahouadji\aff{1} \corresp{\email{l.kahouadji@imperial.ac.uk}}, Cristian R. Constante-Amores\aff{1},
  Gabriel Farah Nor\~{o}es Gon\c{c}alves\aff{1}, Seungwon Shin\aff{2},
  Jalel Chergui\aff{3}, Damir Juric\aff{3}
 \and Omar K. Matar\aff{1}}

\affiliation{\aff{1}Department of Chemical Engineering, Imperial College London,
South Kensington Campus, London SW7 2AZ, United Kingdom\\
\aff{2}Department of Mechanical and System Design Engineering, Hongik University, Seoul 121-791, Republic of Korea\\
\aff{3}Laboratoire d'Informatique pour la M\'ecanique et les Sciences de l'Ing\'enieur (LIMSI), Centre National de la Recherche Scientifique (CNRS), Universit\'e Paris Saclay, B\^at. 507, Rue du Belv\'ed\`ere, Campus Universitaire, 91405 Orsay, France
}
\begin{document}

\maketitle

\begin{abstract}
We study the effect of insoluble surfactants on the wave dynamics of vertically-falling liquid films. We use three-dimensional numerical simulations and employ a hybrid interface-tracking/level-set method, taking into account Marangoni stresses induced by gradients of interfacial surfactant concentration. Our numerical predictions for the evolution of the surfactant-free, three-dimensional wave topology are validated against the experimental work of \cite{Park2003}. The addition of surfactants is found to  influence significantly the development of horseshoe-shaped waves. At low Marangoni numbers, we show that the wave fronts exhibit 
spanwise oscillations before eventually acquiring a quasi two-dimensional shape. In addition, the presence of Marangoni stresses are found to suppress the peaks of the travelling waves and preceding capillary wave structures. At high Marangoni numbers, a near complete rigidification of the interface is observed.
\end{abstract}

\section{Introduction}
The occurrence of falling films in a wide range of industrial and daily-life applications has driven significant interest in the literature and led to comprehensive  reviews 
\citep[see for example][]{Chang1994,Oron1997,Craster2009,Kalliadasis2012}. The complex topological features on the surface of such films have fascinated the scientific community since the ground-breaking experiments by \cite{Kapitza1948}. The desire to isolate the fundamental mechanisms underlying the genesis and development of two-dimensional and three-dimensional waves has led to numerous experimental studies (see, for instance, \cite{Kapitza1948,TailbyS.R.1962,Alekseenko1994,Liu1993,Liu1995,Oron1997,Park2003,Craster2009,Kalliadasis2012} and references therein). These works have uncovered the generation of `families' of waves and the transition from two- to three-dimensional waves. 

Large two-dimensional wave humps dominate the early stages of film development, after which two kinds of secondary transitions help create the spatio-temporal chaos associated with solitary wave structures in falling films \citep{Liu1993, Cheng1995}. \cite{Chang1996} discussed the presence of streamwise two-dimensional secondary instability leading to the coalescence and coarsening of the initially saturated two-dimensional waves. Additionally, a secondary three-dimensional instability initiates the spanwise transformation of the two-dimensional waves \citep{Joo1992}. Two avenues for the transition from two- to three-dimensional waves exist depending on the ratio of the spanwise to streamwise noise level at the inlet: an out-of-phase three-dimensional checkerboard evolution of the two-dimensional wave front is observed at sufficiently large cross-stream noise level \citep{Chang1994a}, and a synchronous horseshoe-shaped modulation of the wave front at weak spanwise noise levels \citep{Liu1995}.

Drawing inspiration from the work of \cite{Liu1995}, \cite{Scheid2006} developed a low-dimensional weighted residual integral boundary layer model to study the two- to three-dimensional transition and found that the herringbone pattern is largely dependent on the initial conditions. Knowledge of the effect of the synchronous spanwise instability has led to the design of the experiments conducted by \cite{Park2003}, who were able to isolate the horseshoe-shaped solitary waves of prescribed spanwise and streamwise wavelength, while bypassing the secondary two-dimensional wave dynamics. Using a similar modulation approach, \cite{Dietze2014} performed numerical simulations of three-dimensional waves, which they then used to provide a comprehensive study of flow structures present within the inertia-dominated large wave hump region as well as within the visco-capillary region. 

Surfactants are surface-active species that act to decrease surface tension, additionally introducing variations of this quantity that give rise to  Marangoni stresses, which drive fluid away from regions of high surfactant concentration (low surface tension). In the context of falling film flows, surfactants have a stabilising effect, a concept perceived in the early studies on this topic \citep{TailbyS.R.1962,Benjamin1964,Whitaker1964}. In recent years, linear stability studies have been a primary tool to study the stabilising effect of surfactants on the falling film wave dynamics \citep{Blyth2004,Shkadov2004,Pereira2008,Karapetsas2013,Karapetsas2014,Bhat2018,Hu2020}. Experimentally, the damping of wave activity has been shown by the works of \cite{Georgantaki2012} and \cite{Georgantaki2016}. More recently, \cite{Bobylev2019} investigated the effect of varying concentration of surfactants and observed that at high concentrations the damping effect is reversed, with wave structures beginning to grow again.

In this paper, we study for the first time the effect of insoluble surfactants on vertical falling films in a three-dimensional, non-linear framework. We implement the same initialisation approach as \cite{Dietze2014} and demonstrate the emergence of oscillatory behaviour in the developing wave fronts for intermediate values of a parameter that characterises the relative significance of the Marangoni stresses. We use our detailed numerical results to elucidate the mechanism underlying this phenomenon. In Section 2, we provide details of the problem formulation along with the numerical technique used to carry out the computation. We discuss our numerical results in Section 3 and include those from a validation study wherein our predictions are benchmarked against the experimental observations of \cite{Park2003}. Finally, concluding remarks are provided in Section 4.

\begin{figure}
    \centering
    \begin{subfigure}[b]{0.49\textwidth}
    \includegraphics[width=1\linewidth]{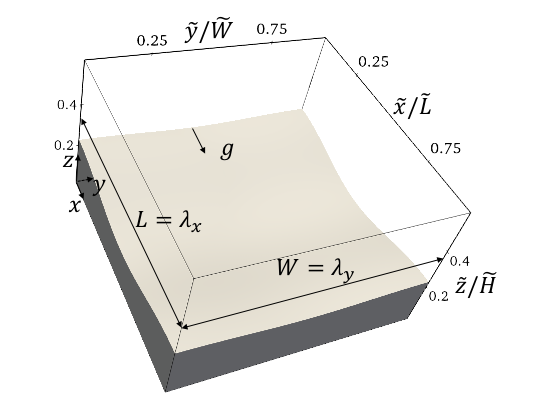}
    \caption{}
    \label{fig:set_up_1}
    \end{subfigure}
                \begin{subfigure}[b]{0.49\textwidth}
    \includegraphics[width=1\linewidth]{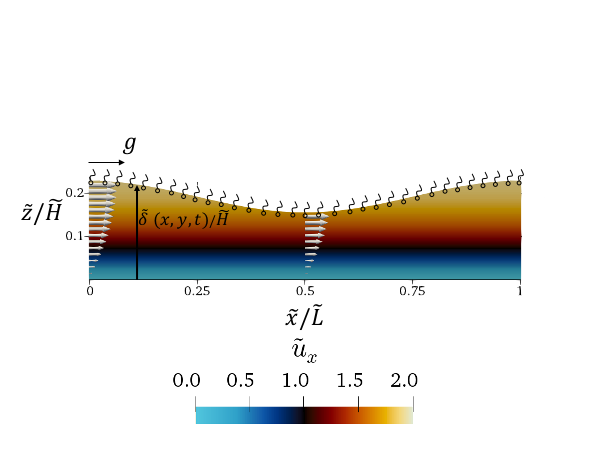}
    \caption{}
    \label{fig:set_up_2}
    \end{subfigure}
    \hfill
    \caption{(a) Initial ($\tilde t=0$) three-dimensional wave profile; (b) schematic representation of the problem in the $x-z$ ($y=0$) plane showing the initial film thickness distribution, $\tilde\delta$, and initial streamwise velocity profile, $\tilde u_x$.}
    \label{fig:set_up} 
\end{figure}

\section{Problem formulation and numerical method}
With the purpose of studying the three-dimensional wave development of vertical falling films in the presence of insoluble surfactant species, we utilise a front-tracking/level set method (also  known  as  the  Level  Contour  Reconstruction  Method as in \cite{Shin2002,Shin_ijnmf_2009,Shin2017}). The continuity and momentum equations are solved in a three-dimensional Cartesian domain, $\textbf{x}=(x,y,z)$, as shown in figure \ref{fig:set_up}, with a coupling for the transport of interfacial surfactant species.
Bulk surfactant transport is not considered in this work. The gas and liquid are assumed to be immiscible, incompressible Newtonian fluids. The full set of dimensional equations for this method can be found in \cite{Shin2018}. To render the equations dimensionless, we use the following scalings:
\begin{equation}
\label{scalings}
\quad \tilde{\textbf{x}}=\frac{\textbf{x}}{\delta_0},
\quad \tilde{\textbf{u}}=\frac{\textbf{u}} {U_0}, 
\quad \tilde{t}=\frac{t}{\delta_0/U_0}, 
\quad \tilde{p}=\frac{p}{\rho_l U_0^2}, 
\quad \tilde{\sigma}=\frac{\sigma}{\sigma_s},
\quad \tilde{\Gamma}=\frac{\Gamma}{\Gamma_\infty},
\end{equation}
\noindent where the tildes designate dimensionless quantities. Here, $t$, $\textbf{u}$, and $p$ denote  time, velocity, and pressure, respectively, and the density of the liquid is $\rho_l$. The mean velocity and thickness of a waveless falling film, as  presented theoretically by \cite{Nusselt1923}, are designated by $U_0$ and $\delta_0$, respectively. The surfactant-free surface tension is $\sigma_s$, while the surface tension coefficient varying with the interfacial surfactant concentration $\Gamma$ is given by an equation of state $\sigma=\sigma(\Gamma)$ given below; the concentration at saturation is given by $\Gamma_\infty$. Using the relations in Eq. (\ref{scalings}), the dimensionless form of the continuity and momentum equations is respectively expressed as:
\begin{equation}
 \nabla \cdot \tilde{\textbf{u}}=0,
\end{equation}
\begin{eqnarray}
 \tilde{\rho}\left(\frac{\partial \tilde{\textbf{u}}}{\partial \tilde{t}}+\tilde{\textbf{u}} \cdot\nabla \tilde{\textbf{u}}\right)  & = & - \nabla \tilde{p}  +  \frac{1}{\Rey}\nabla\cdot  \left [ \tilde{\mu} (\nabla \tilde{\textbf{u}} +\nabla \tilde{\textbf{u}}^T) \right ] +\frac{\textbf{e}_x}{Fr^2} \nonumber\\
 && +\frac{1}{\Web} \int_{\tilde{A}(\tilde{t})} \left(
\tilde{\sigma} \tilde{\kappa} \textbf{n} 
 +  
 \nabla_s  \tilde{\sigma}\right)\delta \left(\tilde{\textbf{x}} - \tilde{\textbf{x}}_{_f}  \right) d\tilde{A} , 
\end{eqnarray}
\noindent where $\tilde\kappa$ denotes the interface curvature, $\nabla_s$ the surface gradient operator, $\textbf{n}$ the outward-pointing unit normal to the interface, and $\textbf{e}_x$ a unit vector in the direction of gravity (i.e. $x$-direction). Here $\tilde{\textbf{x}}_f$ is the parametrization of the time-dependent interface area $\tilde{A}(\tilde{t})$, where $\delta   (\tilde{\textbf{x}}-\tilde{\textbf{x}}_f)$ is the three-dimensional Dirac delta function which vanishes everywhere except at the interface localised at $\tilde{\textbf{x}}=\tilde{\textbf{x}}_f$. The density, $\tilde{\rho}$, and viscosity, $\tilde{\mu}$, are given by
    \begin{equation}
    \label{prop}
    \tilde{\rho}\left( \tilde{\textbf{x}},\tilde{t}\right)=\frac{\rho_g}{\rho_l} + \left(1 -\frac{\rho_g}{\rho_l}\right) H\left( \tilde{\textbf{x}},\tilde{t}\right),~
    \tilde{\mu}\left( \tilde{\textbf{x}},\tilde{t}\right)=\frac{\mu_g}{\mu_l}+ \left(1 -\frac{\mu_g}{\mu_l}\right) H\left( \tilde{\textbf{x}},\tilde{t}\right).
    \end{equation}
\noindent Here, $H\left( \tilde{\textbf{x}},\tilde{t}\right)$ represents a smoothed Heaviside function, which is zero in the gas phase and unity in the liquid phase, while the subscripts $l$ and $g$ designate the individual liquid and gas phases, respectively. Surfactant transport in the context of insoluble surfactants is governed by:
 \begin{equation} 
 \frac{\partial \tilde{\Gamma}}{\partial \tilde{t}}+\nabla_s \cdot (\tilde{\Gamma}\tilde{\textbf{u}}_{\text{t}})=\frac{1}{Pe_s} \nabla^2_s \tilde{\Gamma},
 \end{equation}
 \noindent where $\tilde{\textbf{u}}_{\text{t}}=(\tilde{\textbf{u}}_{\text{s}}\cdot\textbf{t})\textbf{t}$ is the tangential velocity vector in which $\tilde{\textbf{u}}_{\text{s}}$ is the surface velocity and ${\mathbf{t}}$ is the unit tangent to the interface. The dimensionless parameters appearing in these equations are given by
        \begin{equation}\label{dimless}
        \Rey=\frac{\rho_lU_0\delta_0}{\mu_l},~\Web=\frac{\rho_l U^2_0 \delta_0}{\sigma_s},~\Fro=\frac{U_0}{g^{1/2} \delta^{1/2}},~
        ~\Pen_s=\frac{U_0 \delta_0}{D_s},~
        \end{equation}
\noindent
where $\Rey$, $\Web$, and $\Fro$ are the Reynolds, Weber, and Froude numbers, respectively. The gravitational acceleration, assumed to be acting only in $x$ direction, is given by $g$. The surface Peclet number, $\Pen_s$, describes the relative importance of surface diffusion to convection where $D_s$ is the diffusion coefficient. The decrease of  $\sigma$ in relation to $\Gamma$ is modelled using a Langmuir relation \citep{Shin2018}:
\begin{equation} 
\tilde{\sigma}=1 + \beta_s \ln{\left(1 -\tilde{\Gamma}\right)}.
\label{marangoni_eq}
\end{equation}
\noindent The surfactant elasticity parameter is defined as $\beta_s= \Re T\Gamma_\infty/\sigma_s$, where $\Re$ is the ideal gas constant, and $T$ is temperature. Marangoni stresses, which arise from gradients of surface tension, can be expressed in terms of $\tilde\Gamma$ by:
\begin{equation}
\frac{1}{\Web} \nabla_s \tilde{\sigma}\cdot {\mathbf{t}}  \equiv \frac{\tilde{\tau}}{\Web}=-\Mar\frac{1}{(1-\tilde{\Gamma})}\nabla_s\tilde{\Gamma}\cdot {\mathbf{t}},  
\end{equation}
\noindent where $\Mar\equiv \beta_s/\Web= \Rey T\Gamma_\infty/\rho_l U_0^2 \delta_0$ is a Marangoni parameter.

The numerical set-up of the problem closely follows the previous construct of \cite{Dietze2014}. We impose a periodic boundary condition in both streamwise and spanwise directions of the domain shown in figure \ref{fig:set_up_1}. The bottom wall of the domain is treated as a no-slip boundary, whereas a no-penetration boundary condition is prescribed for the top domain boundary. Following the same formulations for the initial film thickness and streamwise velocity profile as \cite{Dietze2014}, we set 
\begin{equation} 
\tilde \delta|_{t=0}=\tilde {\bar \delta}\left[1+0.2\cos\left(\frac{2\pi \tilde x}{\tilde{\lambda}_x}\right)+0.05\cos\left(\frac{2\pi \tilde y}{\tilde{\lambda}_y}\right)\right],~
\tilde u_l=\frac{\Rey}{\Fro^2}\left(\tilde \delta\tilde z-0.5\tilde{z}^2\right),
\label{initial_cond}
\end{equation}
\noindent where $\tilde{\lambda}_x$ and $\tilde{\lambda}_y$ are the dimensionless domain length and width, respectively, and are motivated by the experimental set-up of \cite{Park2003}. In their work, \cite{Dietze2014} acknowledge the importance of $\tilde{\bar\delta}$ as a control parameter for liquid volume (since $\tilde{V}_l=\tilde{\bar{\delta}} \tilde{\lambda}_x\tilde{\lambda}_y$), Reynolds number, and streamwise wave frequency in numerical cases where periodic boundary conditions are employed. In this work, we have set up the base surfactant-free validation case using the height, $\tilde H$, and the mean film thickness, $\tilde{\bar\delta}$, estimated in the work of \cite{Dietze2014}. As a result, our domain size has the following dimensions: $0.025 \times 0.02 \times 0.0012~\rm{m}^3$. 

The targeted flow conditions for the surfactant-free base case are from \cite{Dietze2014} $\Rey=59.3$, $\Web=0.159$, $\Fro=4.45$, and $f=17Hz$. The selected fluid properties are representative of an air-water system at $25^{\circ}$C: $\rho_l/\rho_g=841.41$, $\mu_l/\mu_g=48.22$, and $\sigma_s=0.072$ N/m. For surfactant-laden cases, the initial surfactant distribution is uniform across the interface (e.g. $\tilde{\Gamma}=0.1$) as shown in figure \ref{fig:set_up_2}. The surface Peclet number also remains unchanged from $\Pen_s=10$ for all surfactant simulations. The Marangoni parameter is the key focus of this work and is varied in the range $0.63-1.88$. For the chosen initial surfactant concentration and Marangoni parameters, the maximum overall surface tension reduction is $3.2\%$, which is sufficiently low to allow us to analyse the effect of Marangoni stresses without the need for running additional simulations where $\tilde\tau=0$. To assess the grid dependence of the results we have used two uniform Cartesian grids for the surfactant-free base case: $M_1=768\times576\times64$  and $M_2=1536\times1152\times128$. All surfactant simulations were run using mesh $M_1$, unless stated otherwise in the text.
\section{Results}
\begin{figure}
\begin{center}
    \begin{subfigure}[b]{0.19\textwidth}
    \begin{center}
     \includegraphics[width=0.94\linewidth]{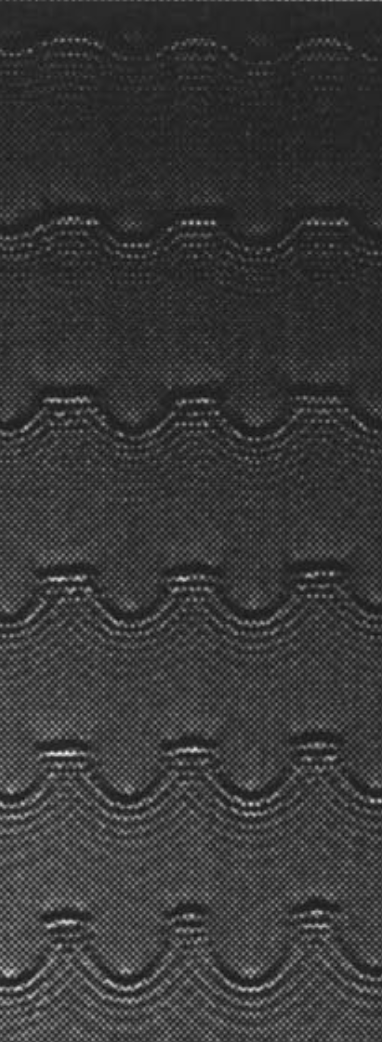}\\
       \caption{}
       \label{fig2:PN}
     \end{center}
    \end{subfigure}
    \hspace{-0.2cm}
    \begin{subfigure}[b]{0.19\textwidth}
    \begin{center}
    \includegraphics[width=1\linewidth]{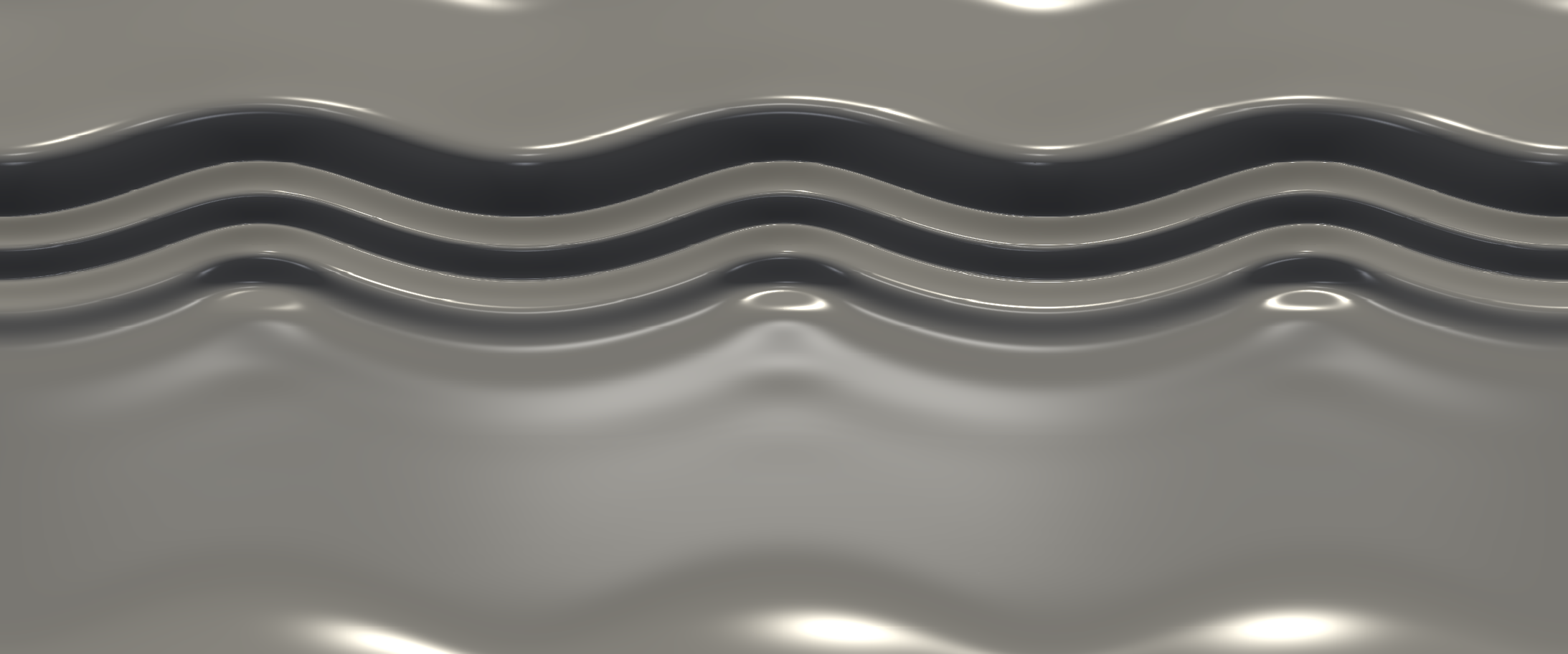}  \\ 
    \includegraphics[width=1\linewidth]{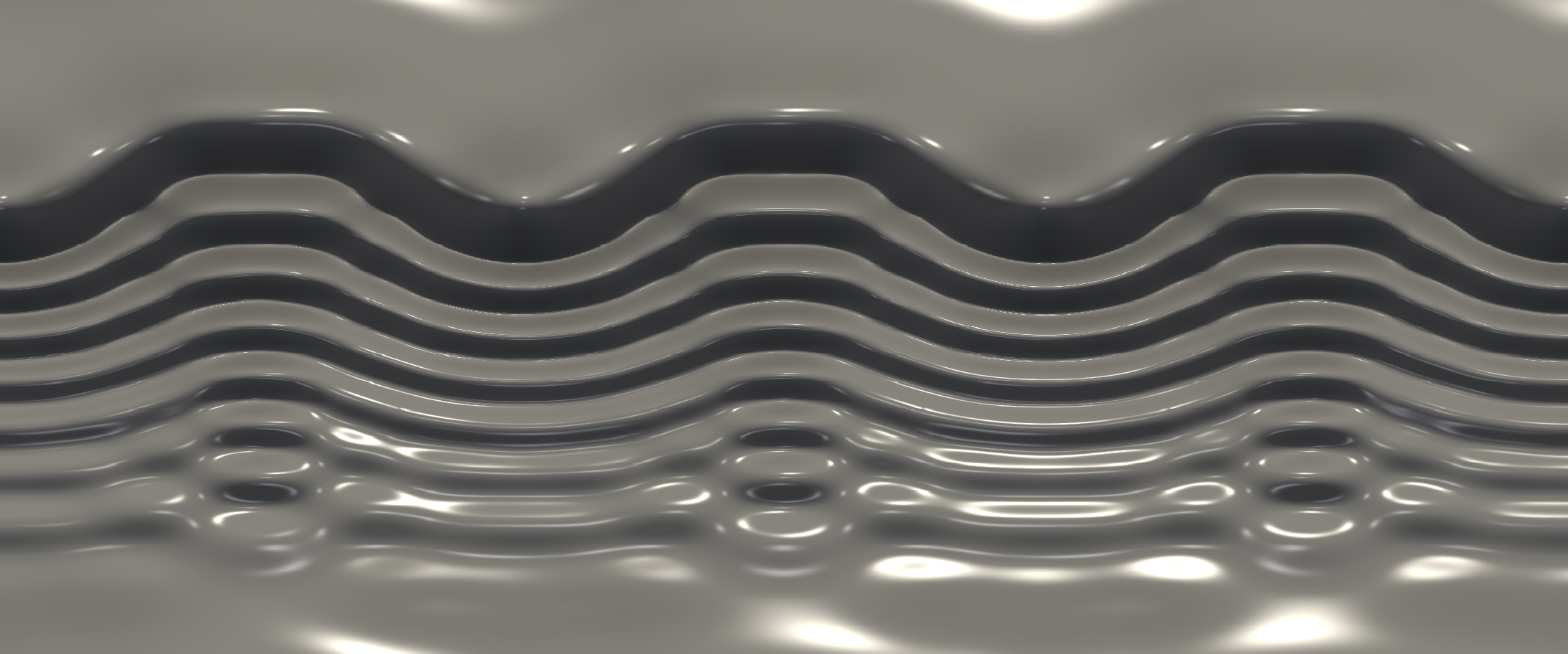}  \\ 
    \includegraphics[width=1\linewidth]{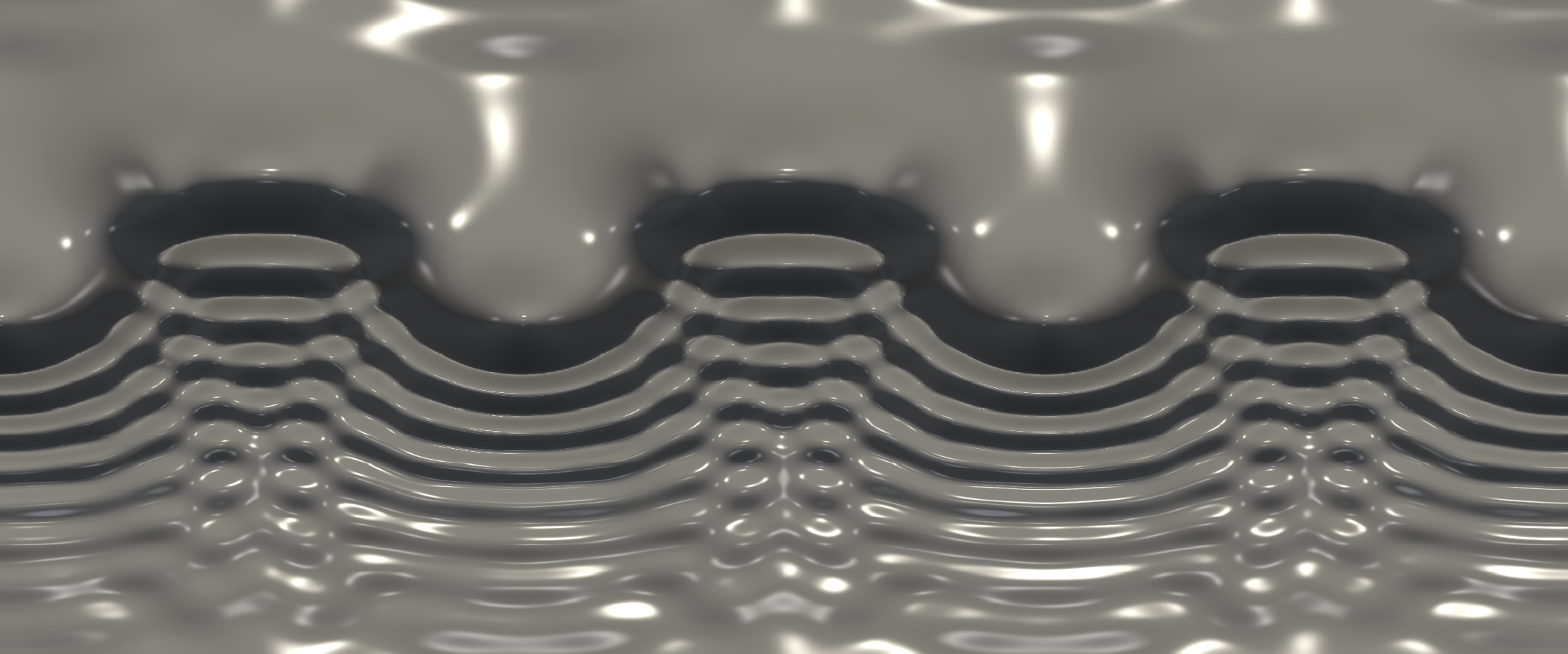}  \\ 
    \includegraphics[width=1\linewidth]{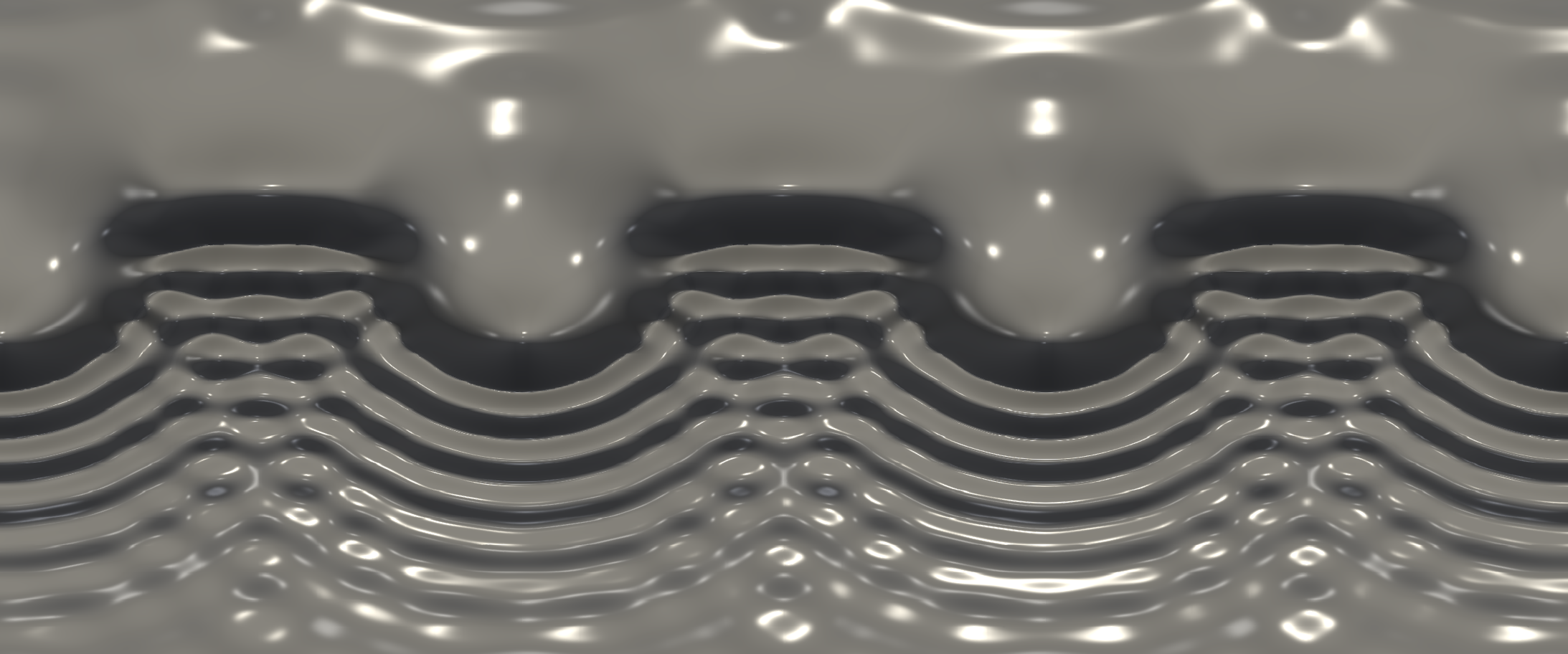}  \\
    \includegraphics[width=1\linewidth]{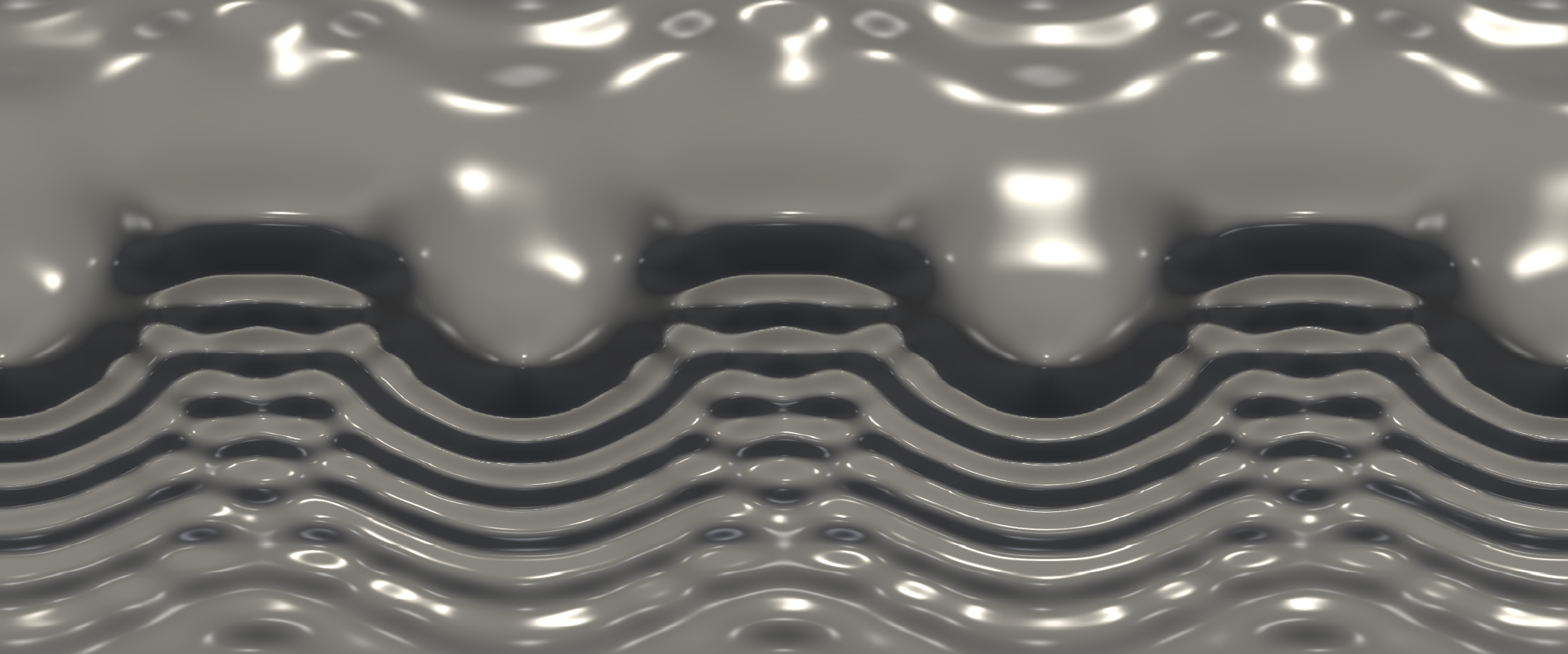}  \\
    \includegraphics[width=1\linewidth]{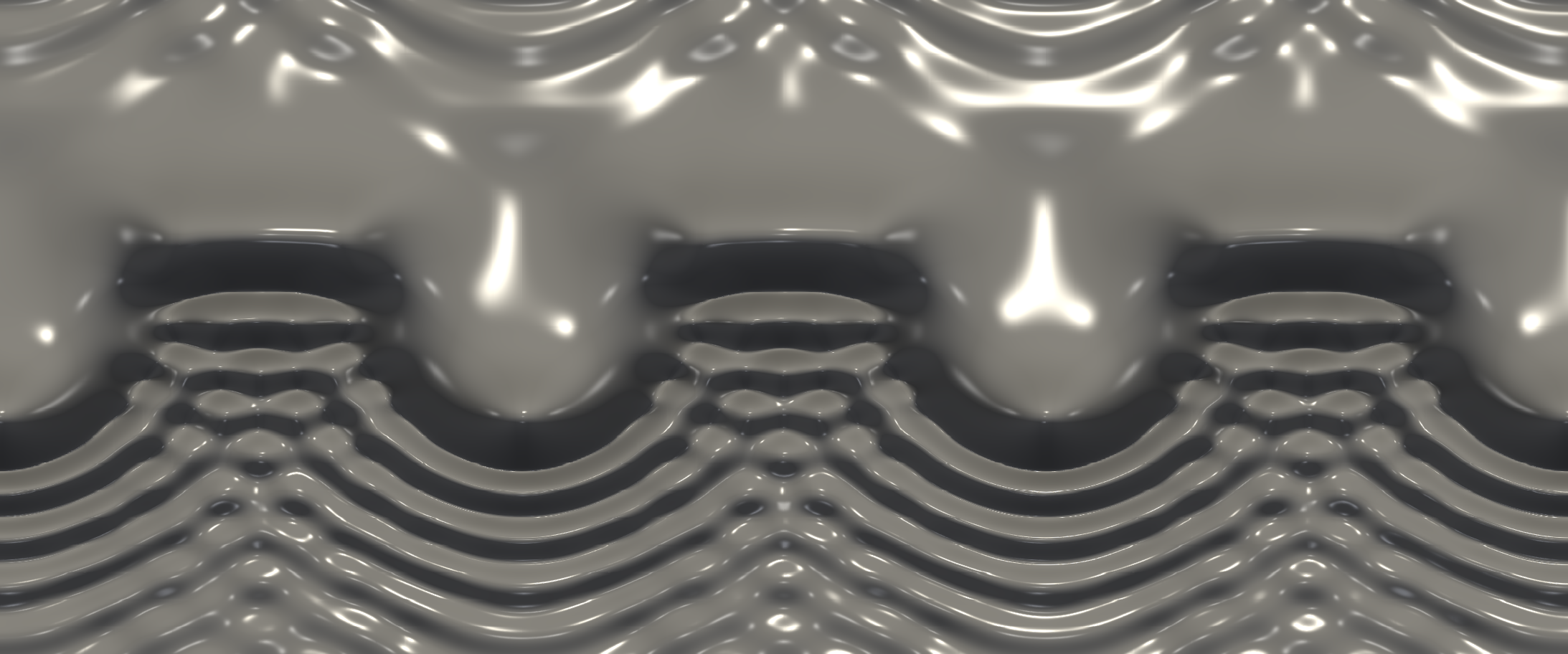}  \\    
        \caption{}
        \label{fig2:clean}
     \end{center}
    \end{subfigure}
     \begin{subfigure}[b]{0.19\textwidth}
    \begin{center}
    \includegraphics[width=1\linewidth]{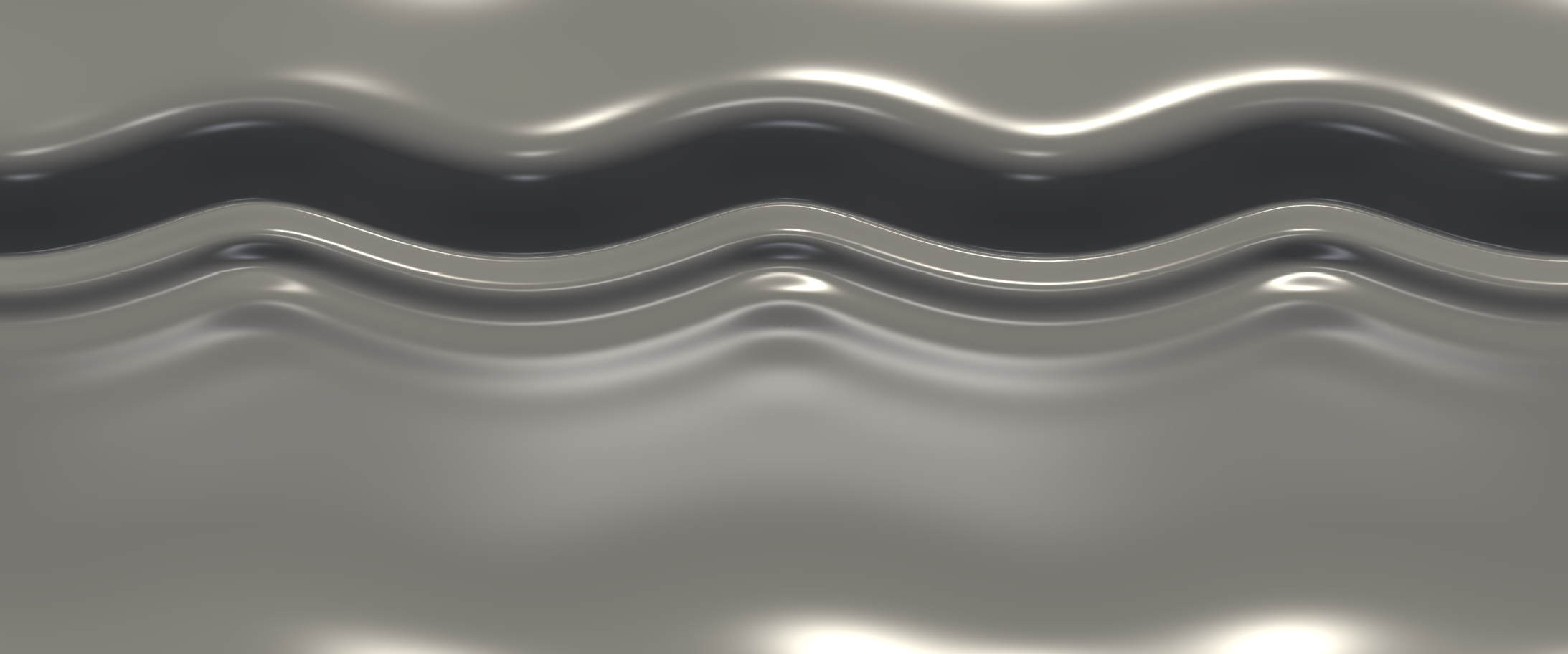}  \\ 
    \includegraphics[width=1\linewidth]{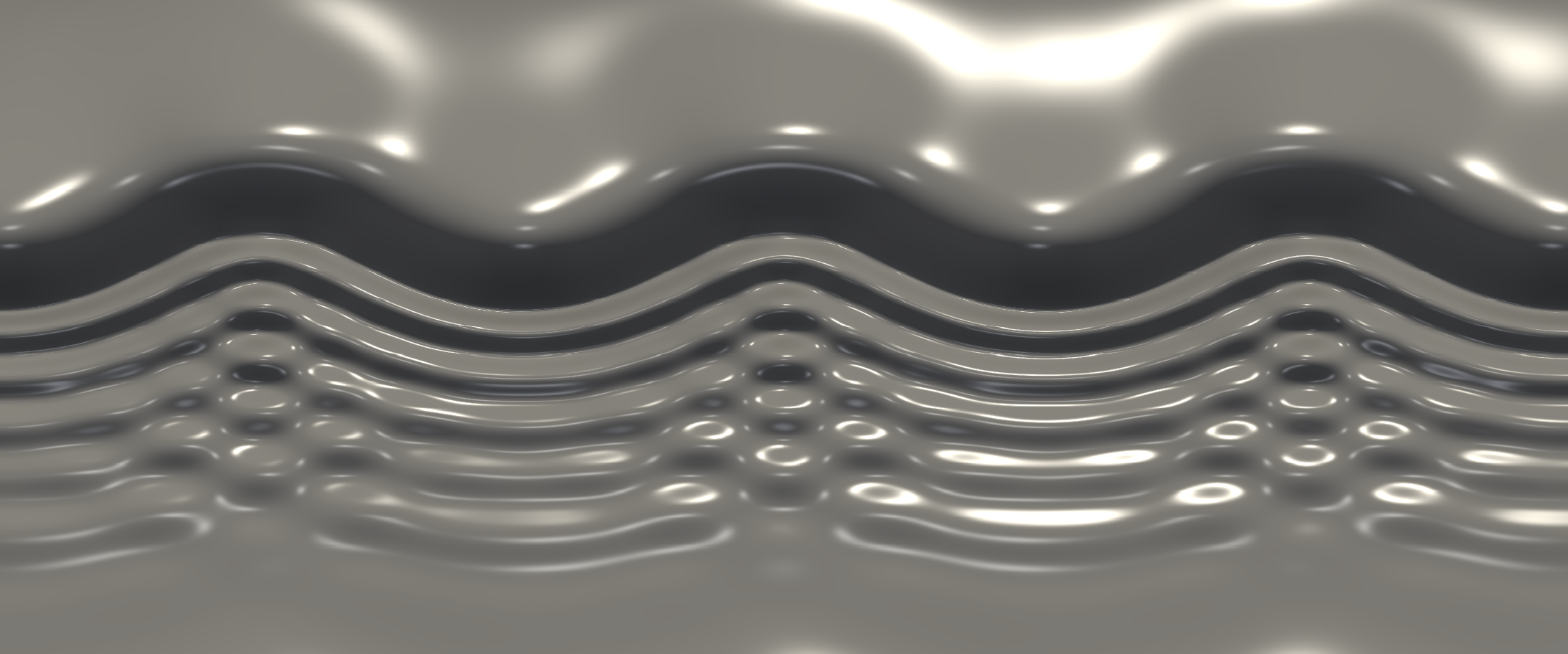}  \\ 
    \includegraphics[width=1\linewidth]{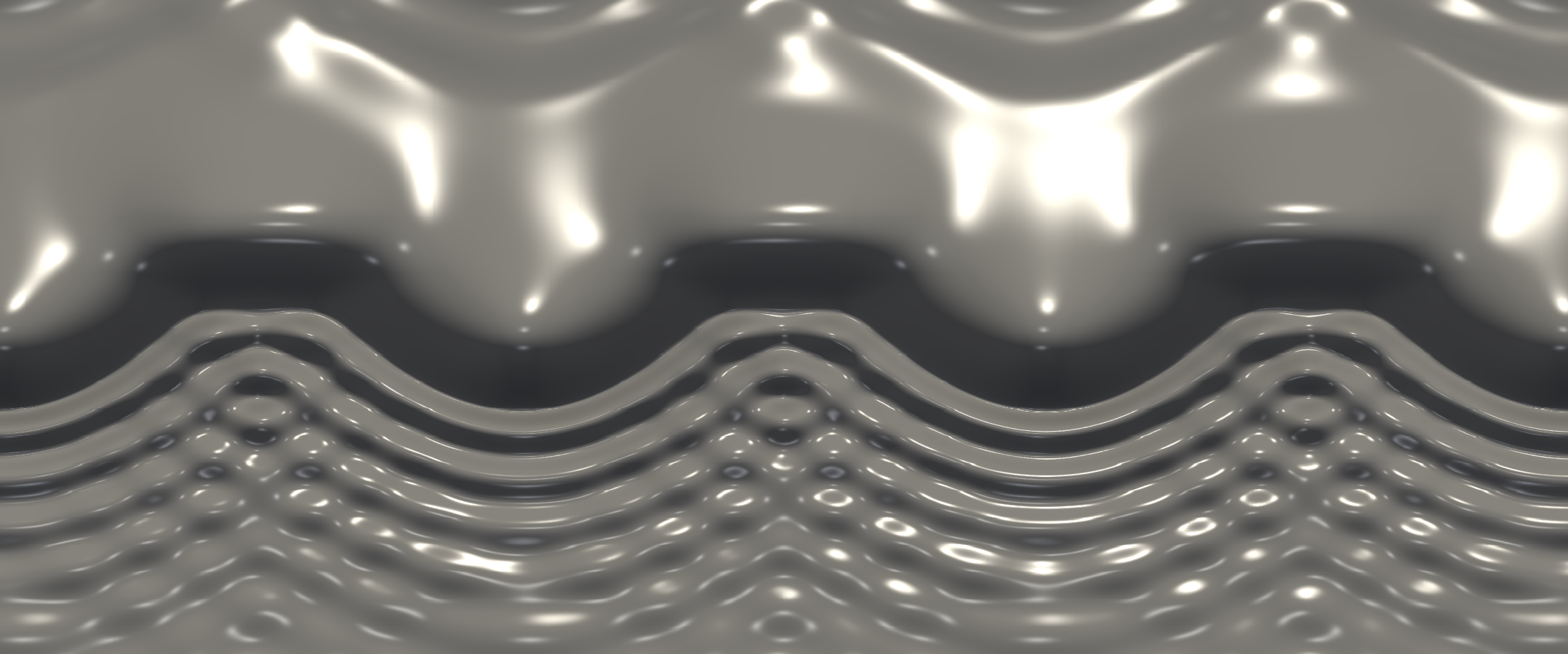}  \\ 
    \includegraphics[width=1\linewidth]{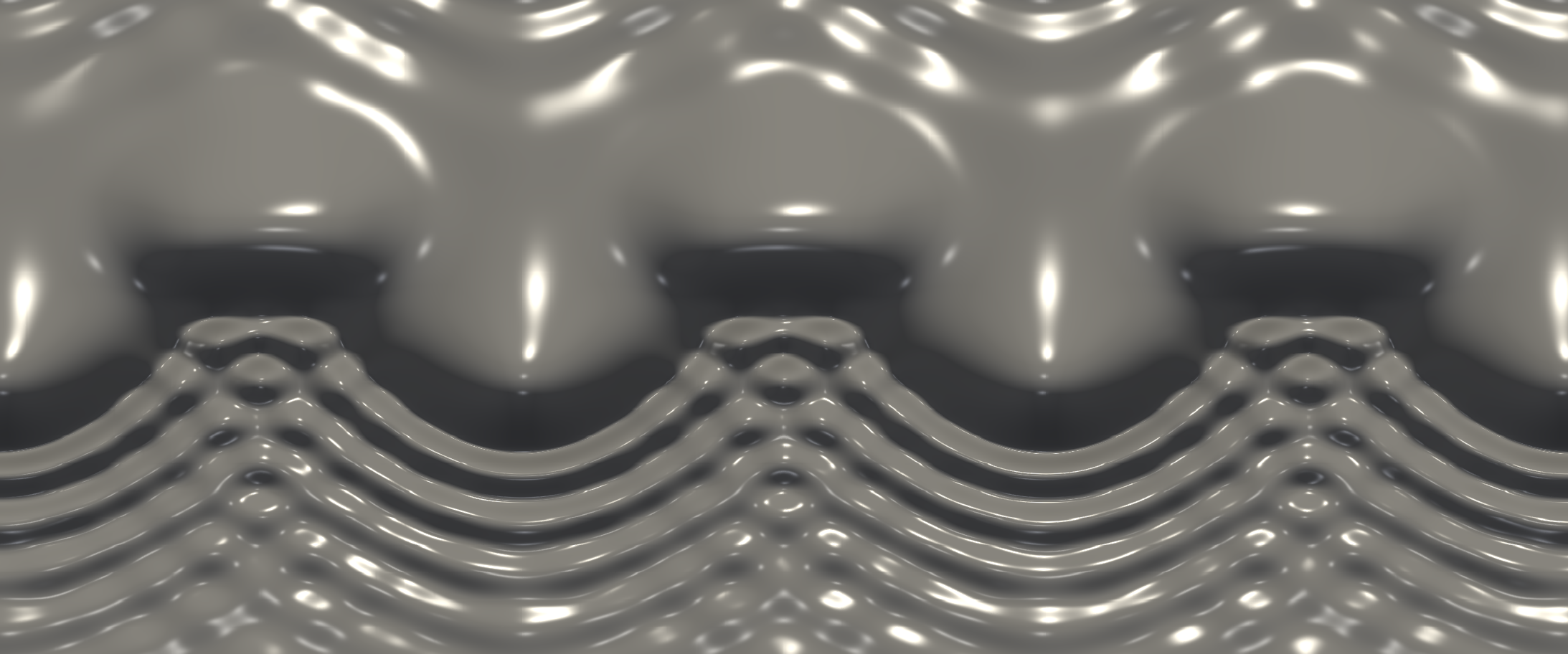}  \\ 
    \includegraphics[width=1\linewidth]{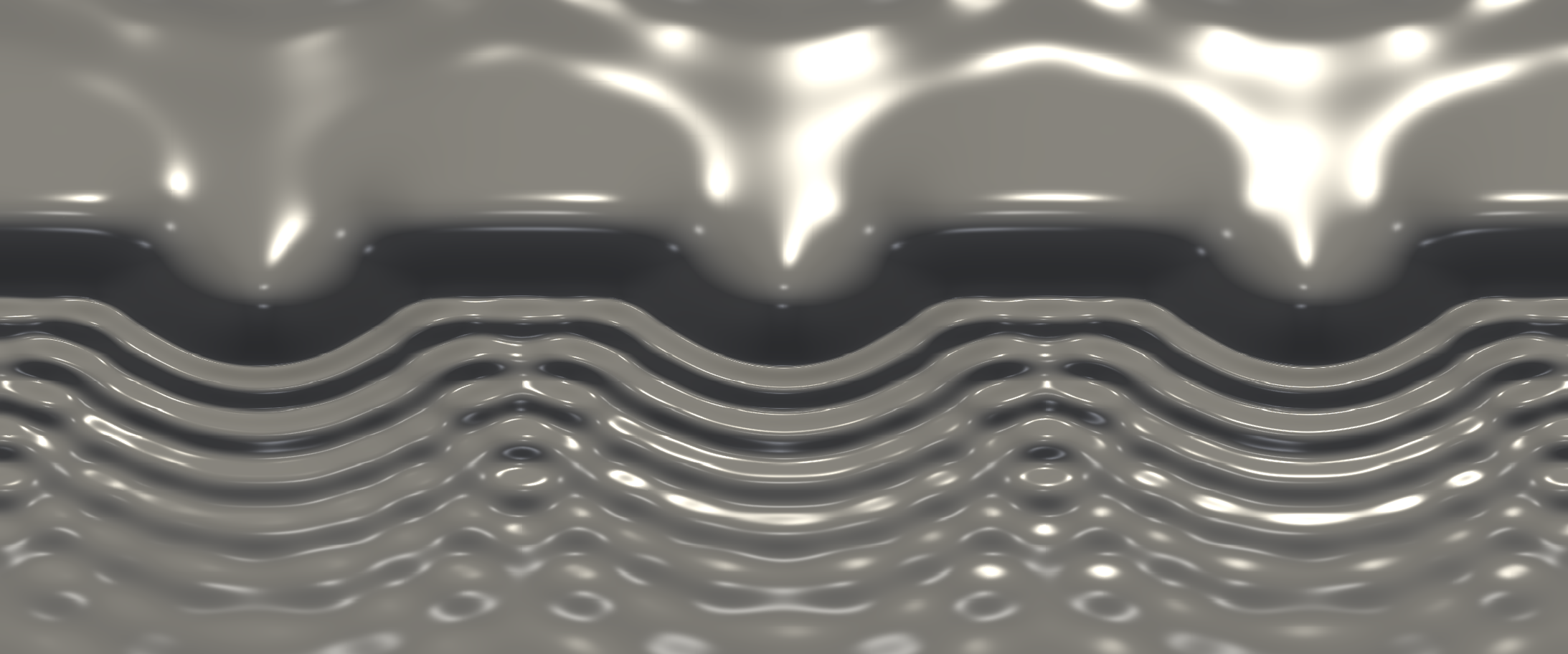}  \\
    \includegraphics[width=1\linewidth]{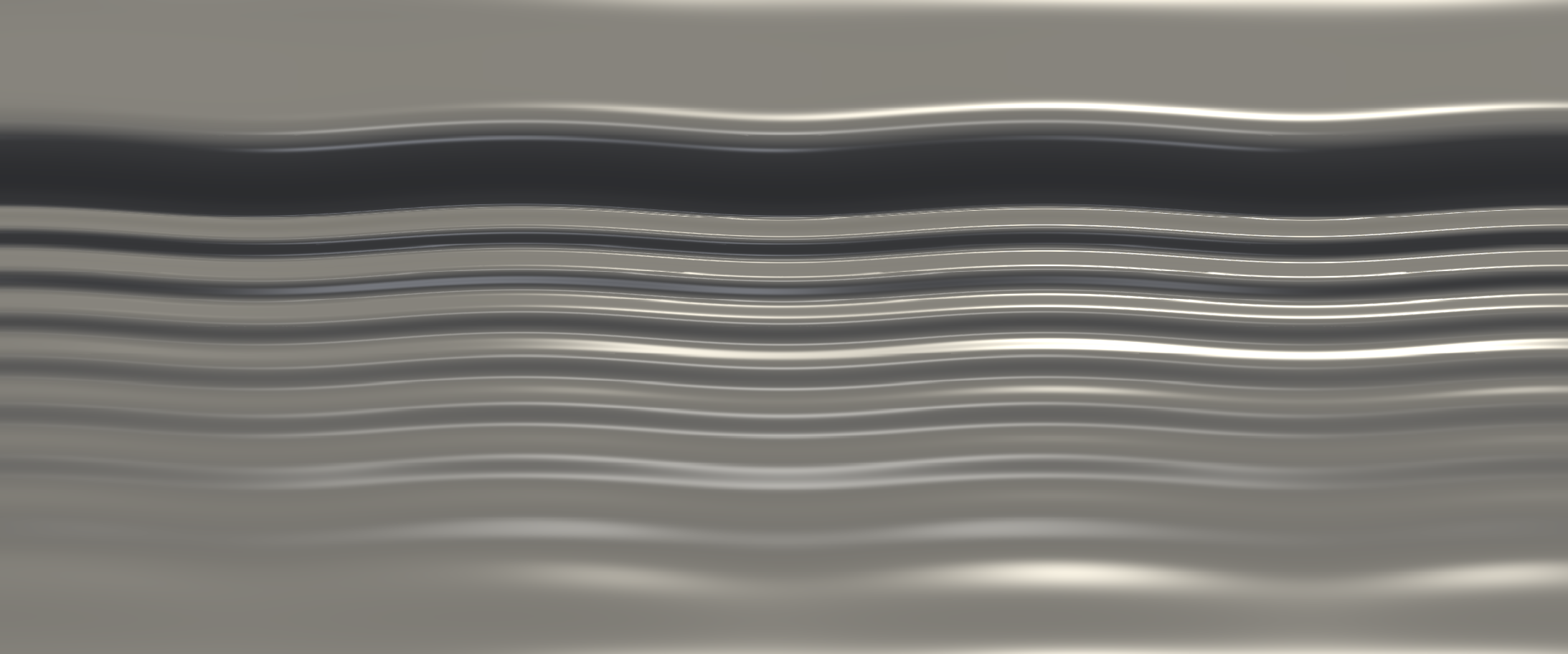}  \\
          \caption{}
          \label{fig2:beta01}
     \end{center}
    \end{subfigure}
    \begin{subfigure}[b]{0.19\textwidth}
    \begin{center}
    \includegraphics[width=1\linewidth]{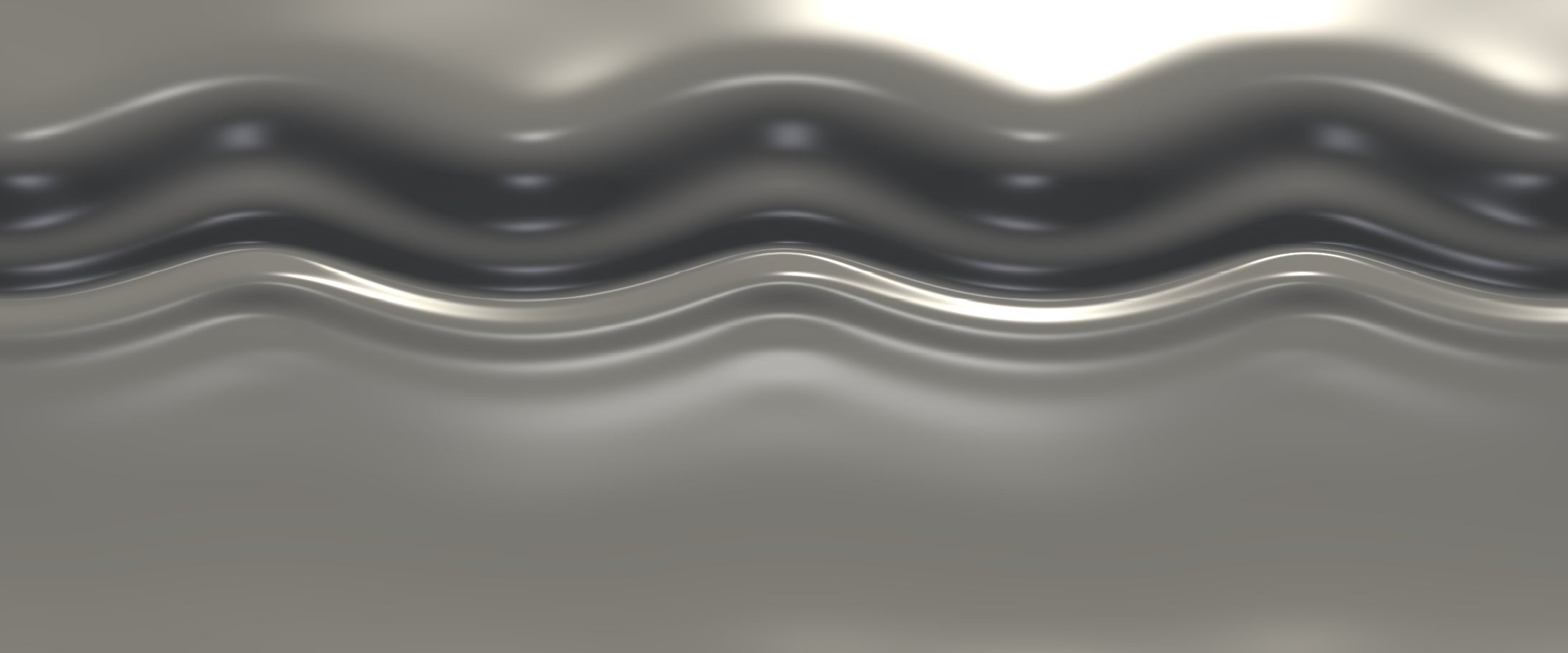}  \\ 
    \includegraphics[width=1\linewidth]{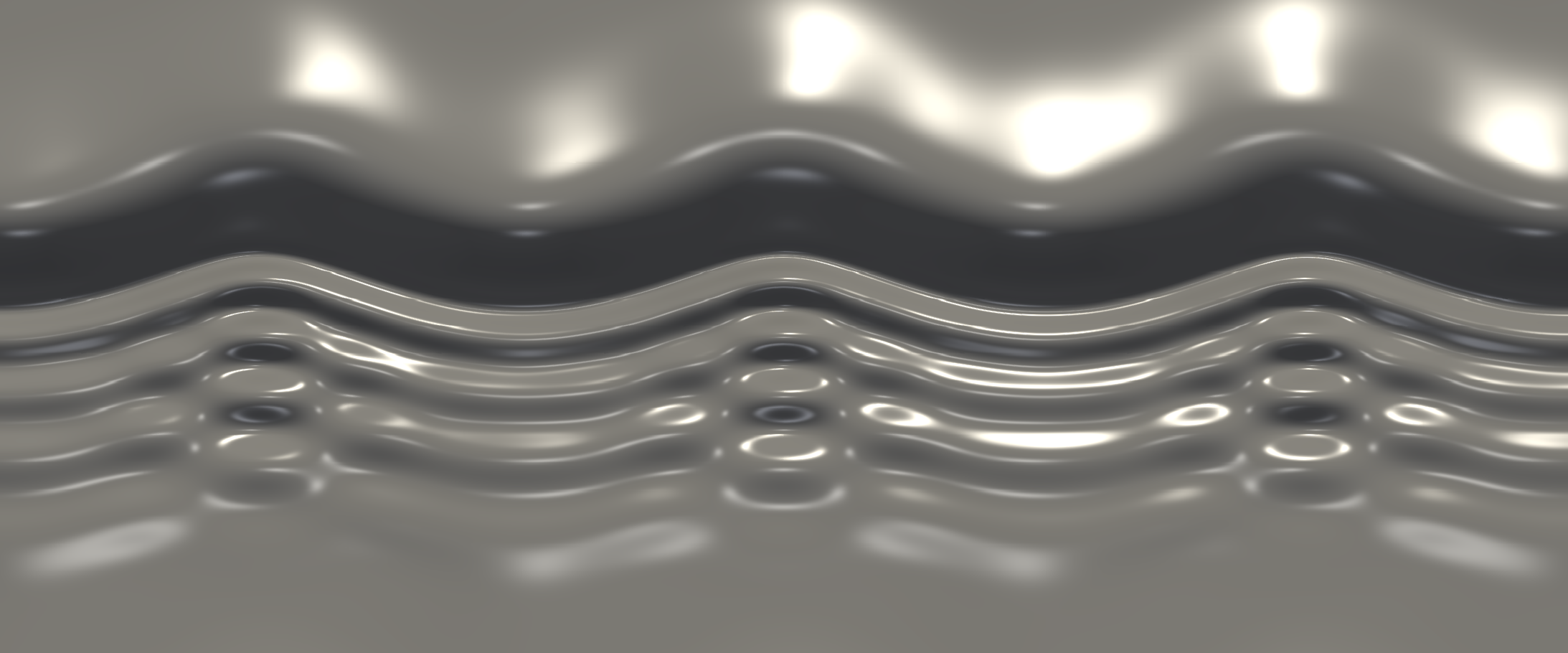}  \\ 
    \includegraphics[width=1\linewidth]{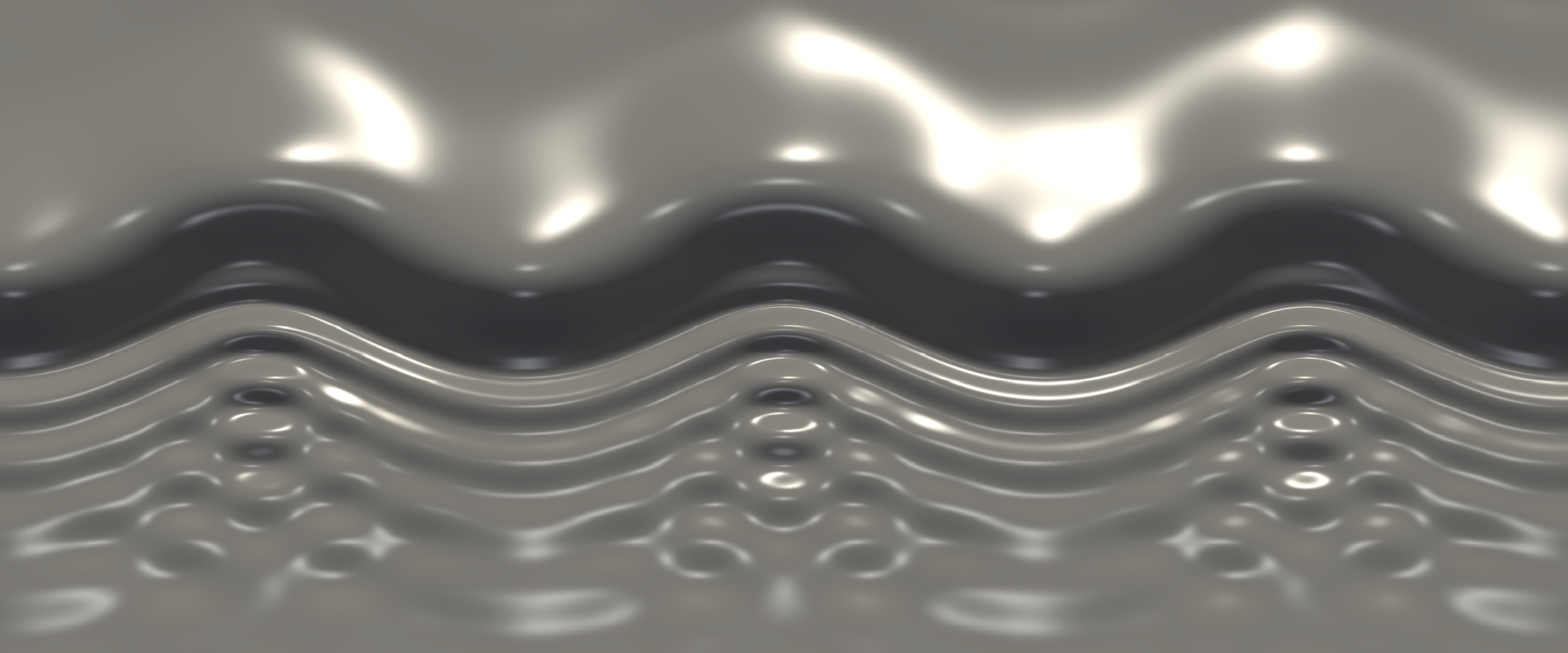}  \\ 
    \includegraphics[width=1\linewidth]{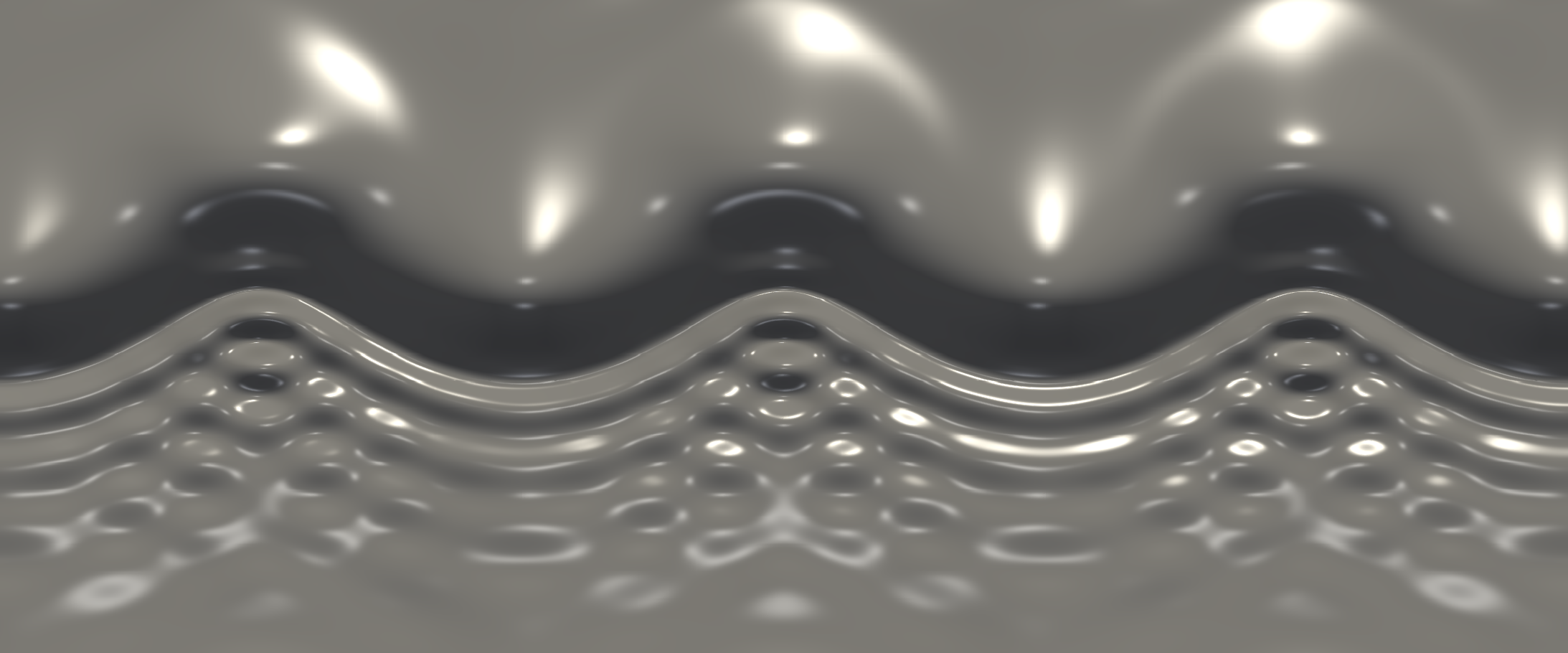}  \\ 
    \includegraphics[width=1\linewidth]{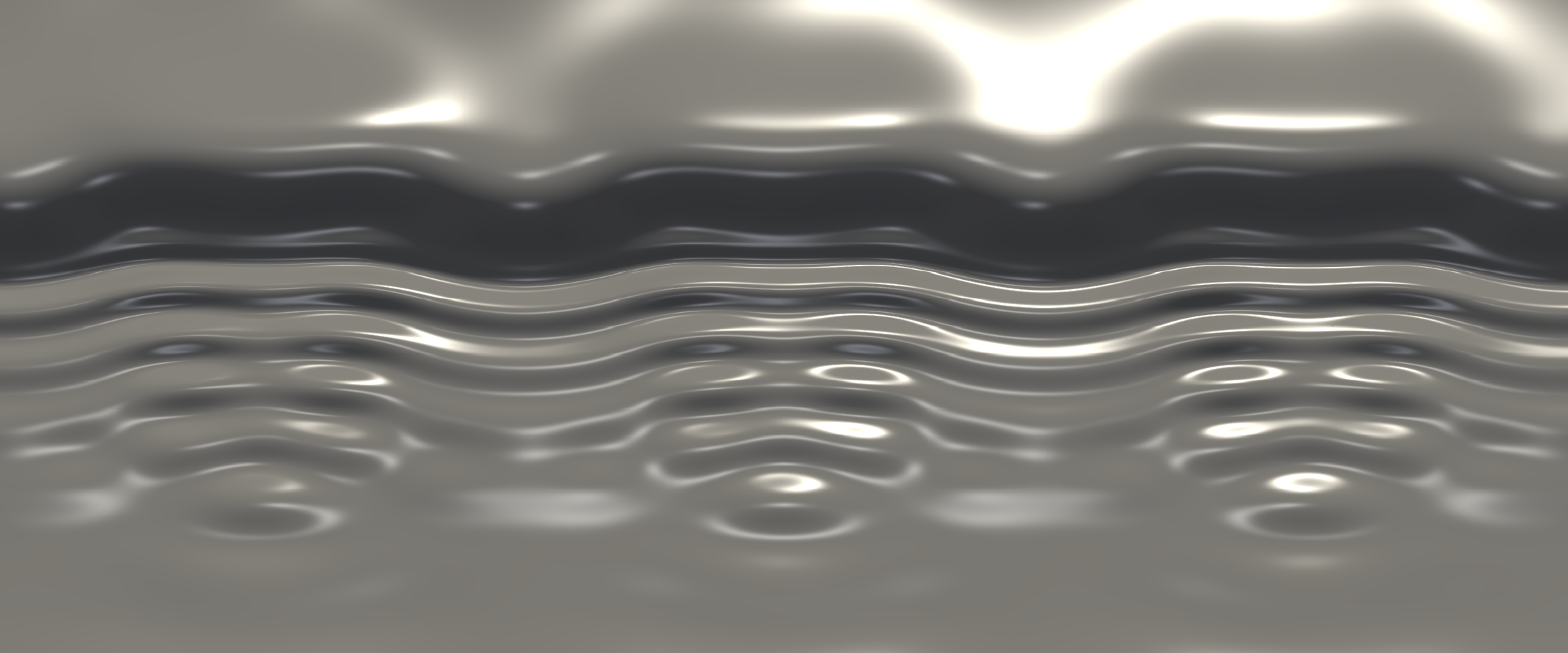}  \\
    \includegraphics[width=1\linewidth]{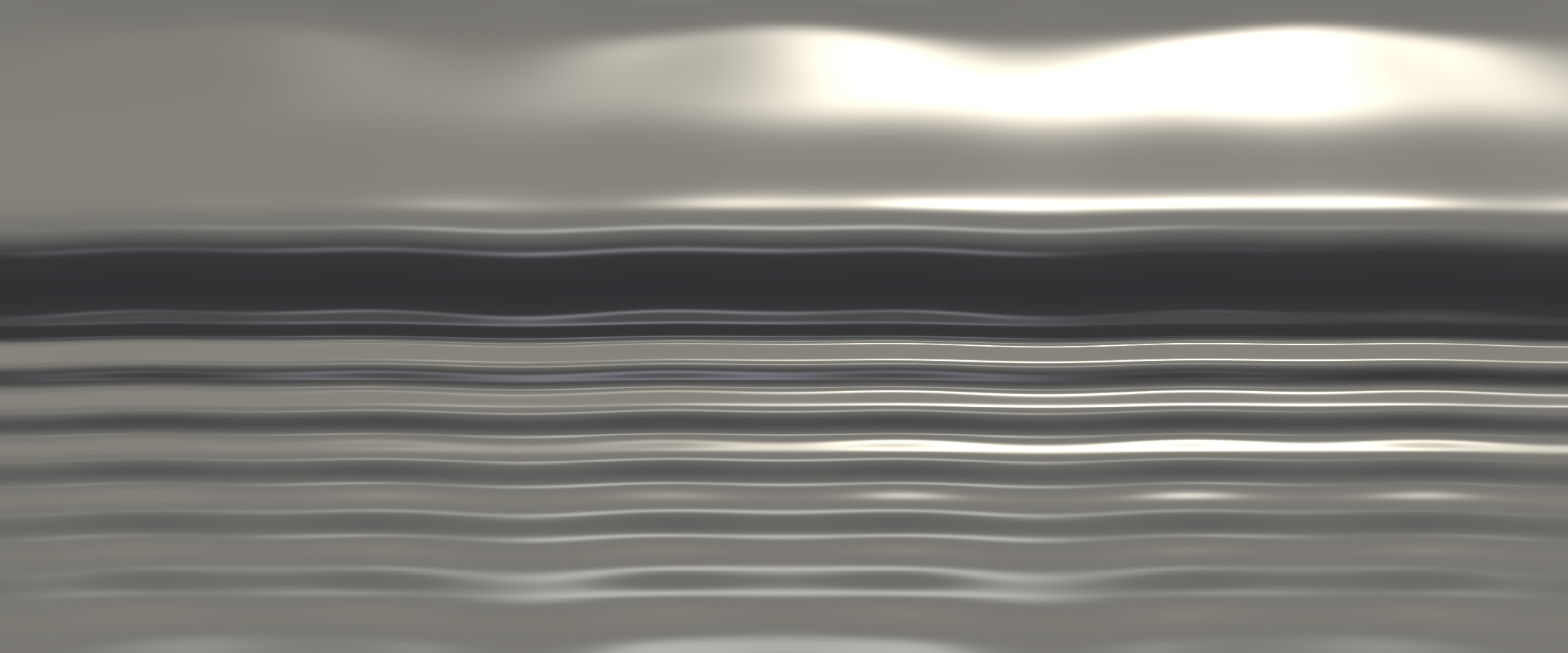}  \\
         \caption{}
         \label{fig2:beta02}
     \end{center}
    \end{subfigure}
    \begin{subfigure}[b]{0.19\textwidth}
    \begin{center}
    \includegraphics[width=1\linewidth]{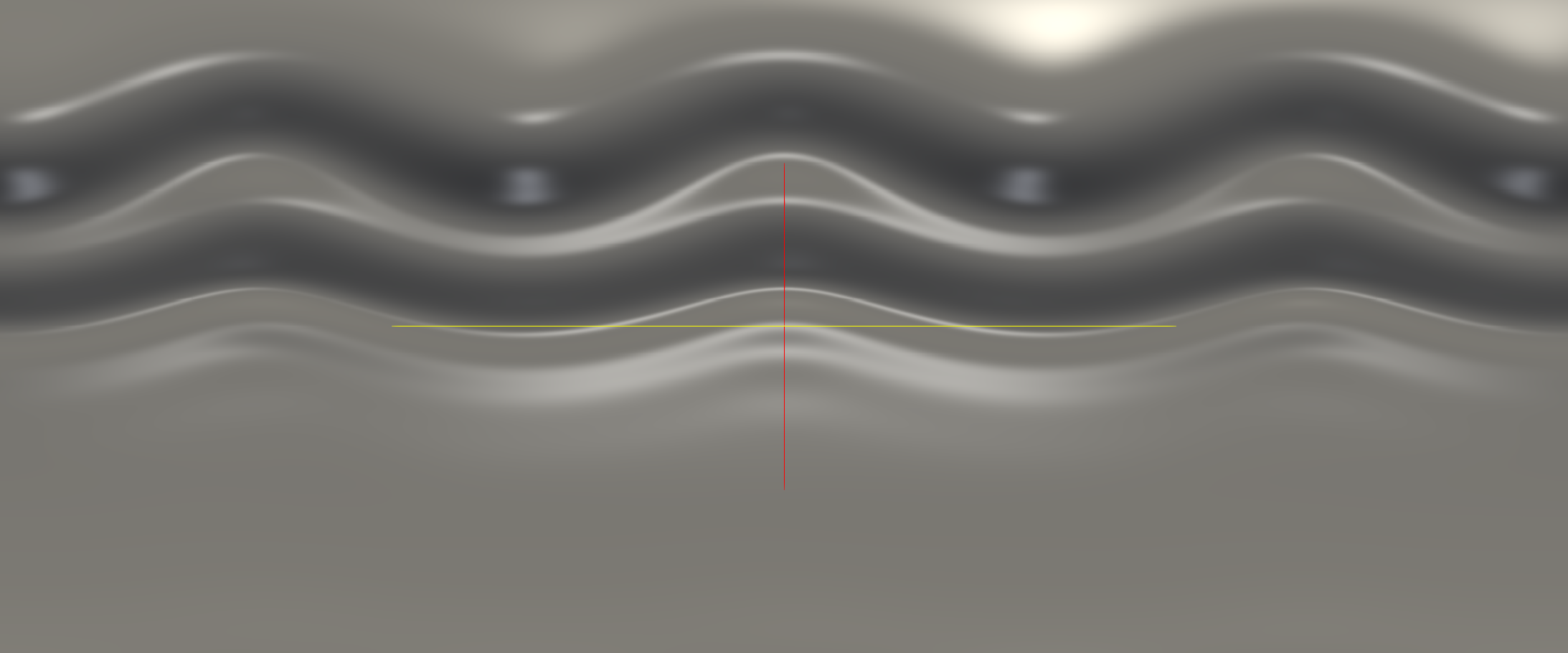}  \\ 
    \includegraphics[width=1\linewidth]{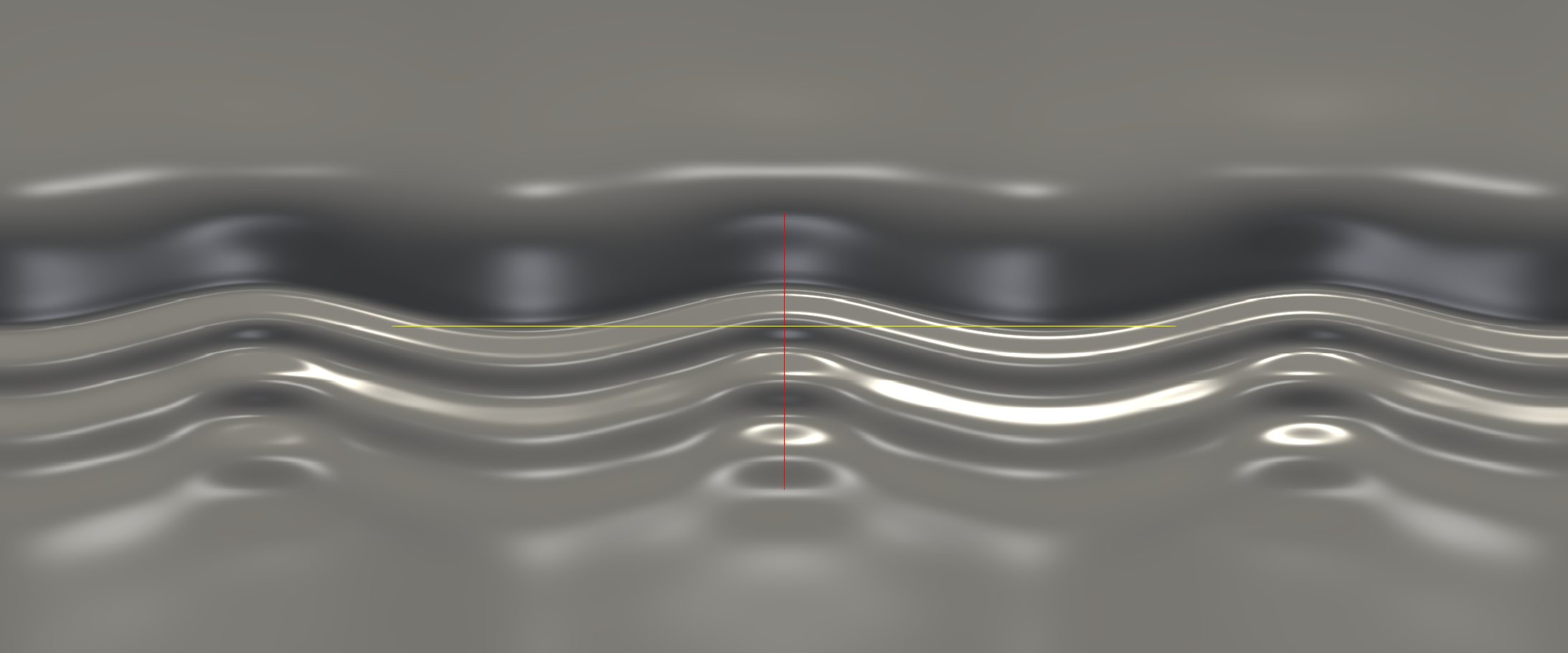}  \\ 
    \includegraphics[width=1\linewidth]{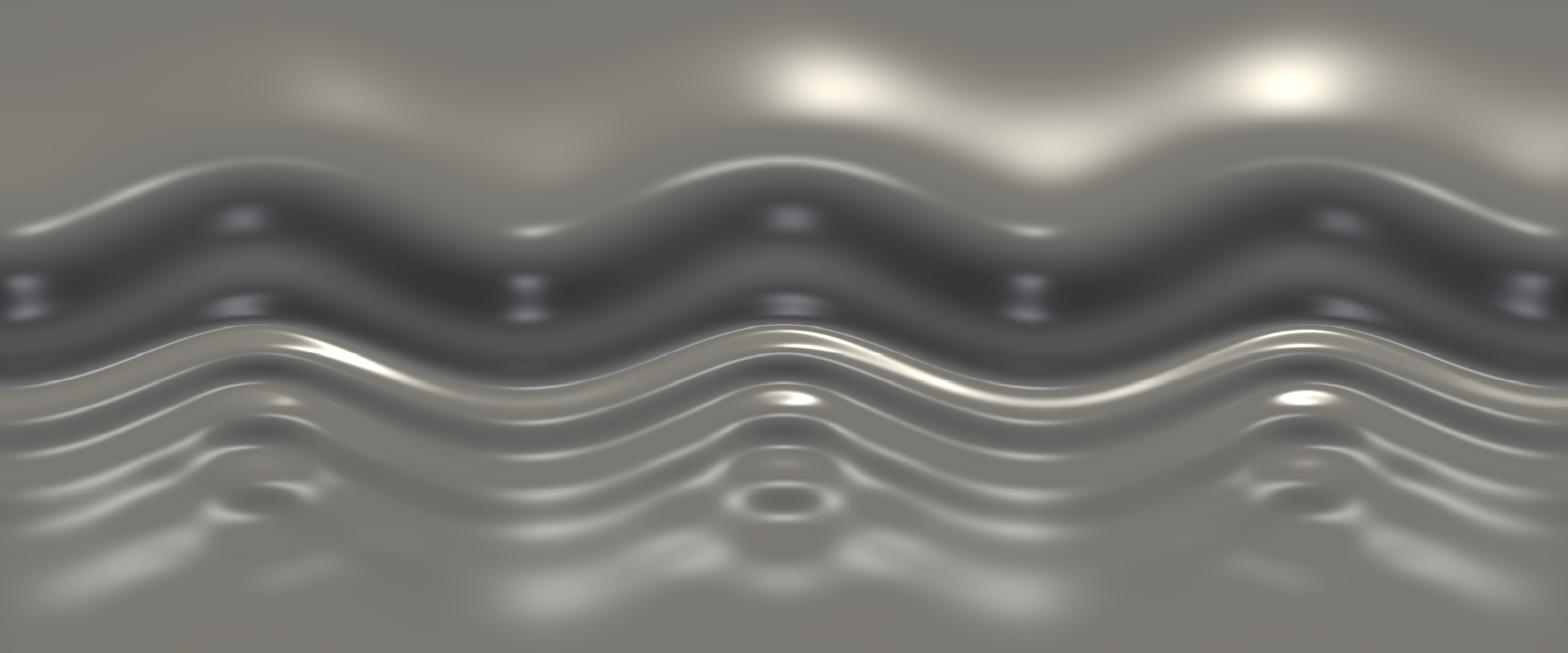}  \\ 
    \includegraphics[width=1\linewidth]{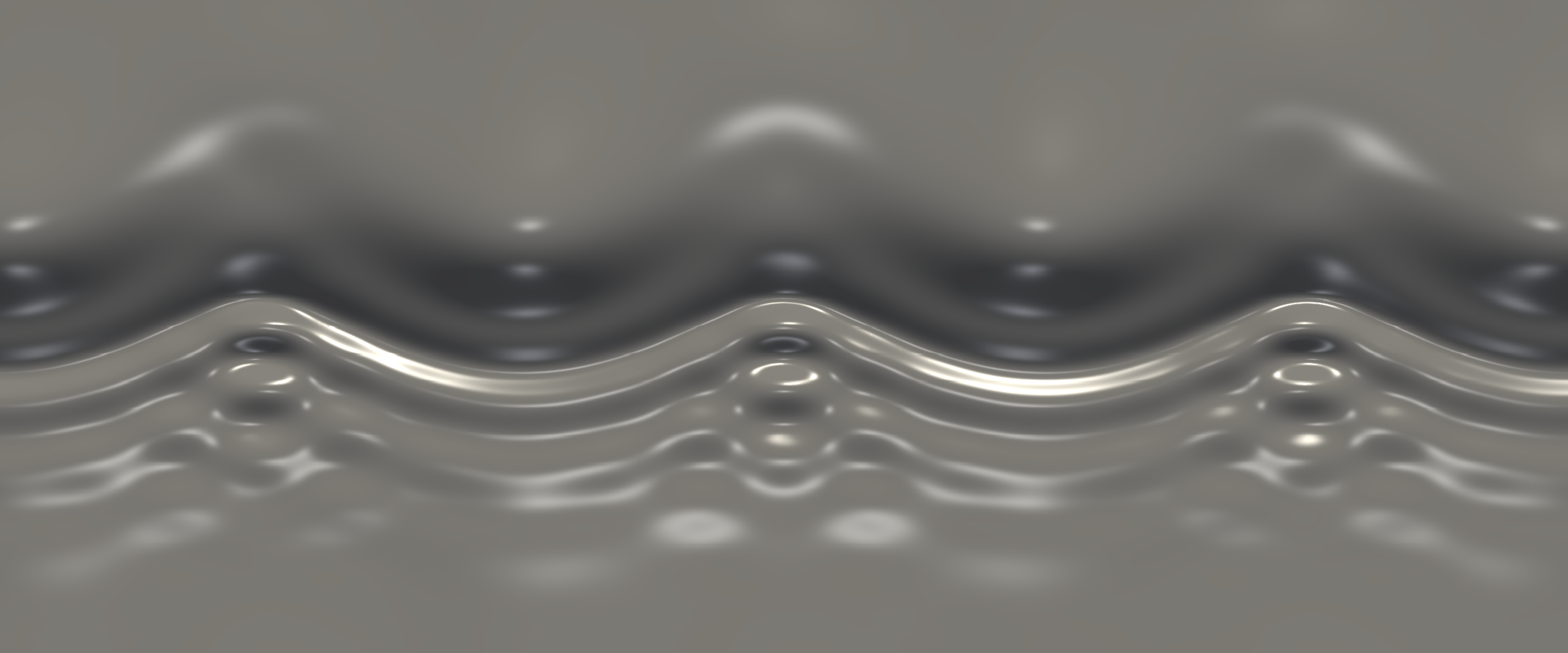}  \\ 
    \includegraphics[width=1\linewidth]{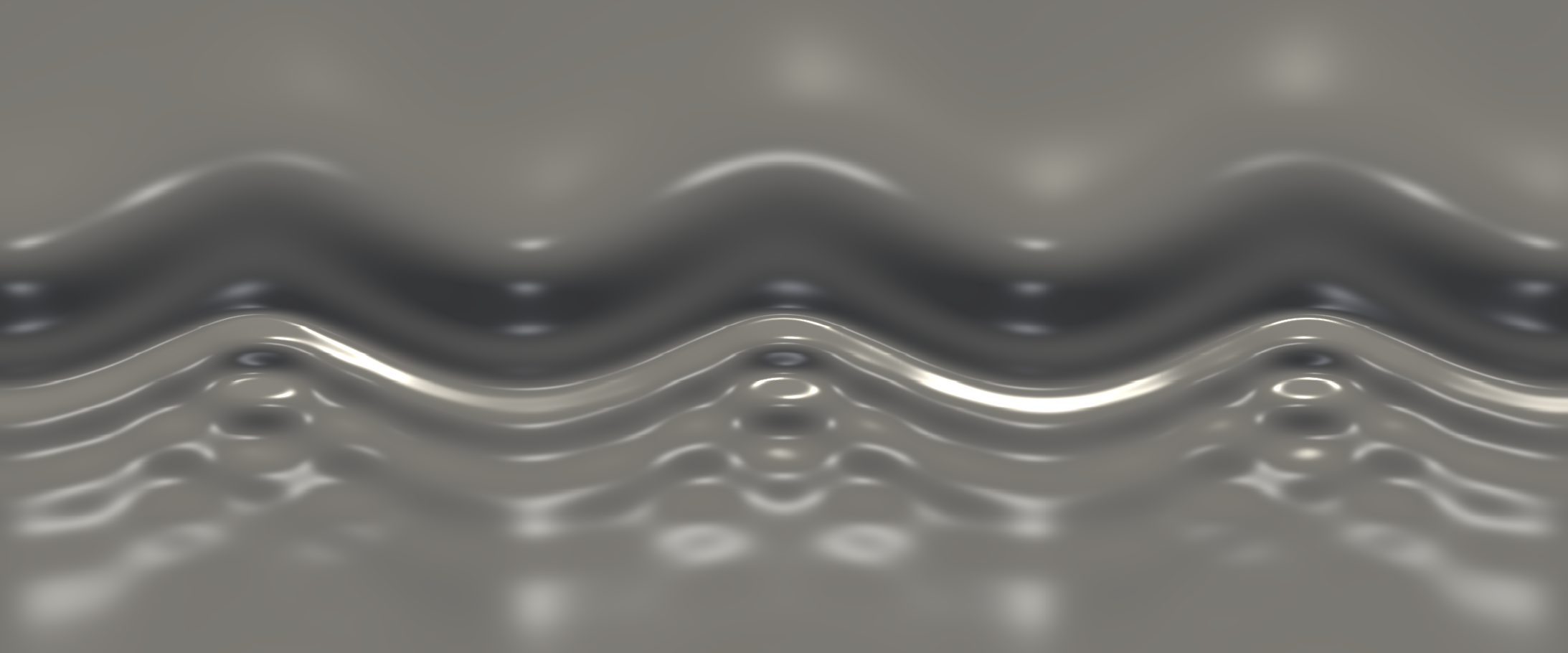}  \\
    \includegraphics[width=1\linewidth]{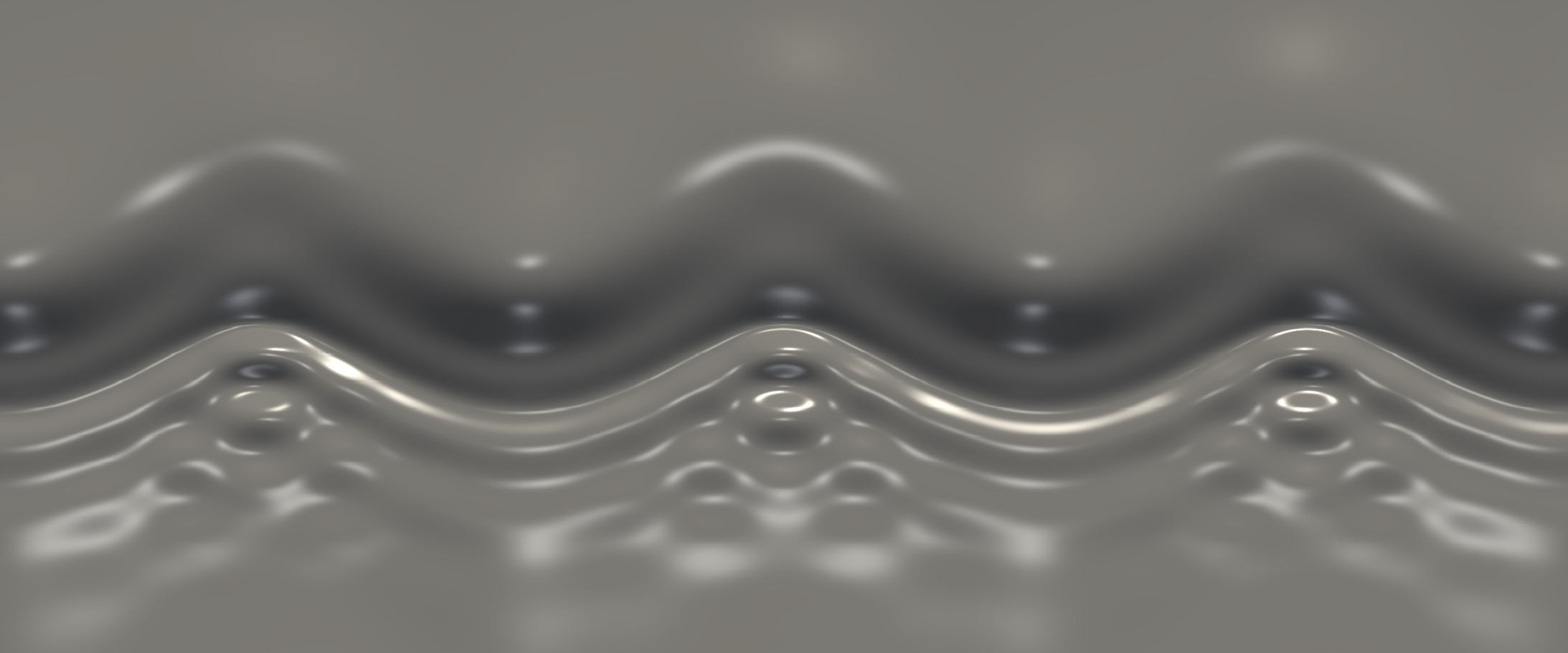}  \\
          \caption{}
          \label{fig2:beta03}
     \end{center}
    \end{subfigure}
     \end{center}
\caption{(a) Experimental snapshot from \cite{Park2003} and simulation results for surfactant-free, (b), and surfactant-laden falling films, (c)-(e). The times associated with panels (b)-(e) are: (b) $\tilde{t}=(60,~116,~170,~223,~277,~330)$; (c) $\Mar=0.63$ at $\tilde{t}=(60,~116,~170,~223,~625,~1111)$; (d) $\Mar=1.25$ at $\tilde{t}=(60,~116,~170,~223,~636,~794)$; (e) $\Mar=1.88$ at $\tilde{t}=(60,~116,~170,~223,~277,~330)$. The surfactant-free and surfactant-laden Reynolds, Weber, and Froude numbers are $59.3$, $0.159$, and $4.45$, respectively. The surface Peclet number is kept constant for all surfactant cases at $Pe_s=10$.}
\label{fig2:fig2}
\end{figure}
We start the discussion of the results by comparing our numerical predictions for the surfactant-free case against the experimental results of \cite{Park2003} (see figure \ref{fig2:PN} and figure \ref{fig2:clean}). The numerical results are given as snapshots of the wave fronts at  times corresponding to subsequent periods, whereas the experimental shadowgraph shows the evolution of the waves in the streamwise direction. During the initial stages, the dynamics are dominated by small-amplitude sinusoidal undulations that develop into well-defined `horseshoe' shapes separated by flat regions.  We also capture the development of the capillary waves, which precede the horseshoes, and their interactions.  It is evident from figure \ref{fig2:clean} that our numerical technique permits the simulation of the wave evolution with good accuracy. The disparity with the experimental observations of \cite{Park2003} in terms of over-estimation of the horseshoe spacing is similar to that in the work of \cite{Dietze2014}; these authors argued that this is an artefact of the imposed constant spanwise wavelength, $\tilde{\lambda}_y$, whereas in the experiments this wavelength varies as the wave fronts travel downstream. In addition to the validation against experimental data, we have also carried out mesh-sensitivity studies. The two key monitored quantities, the kinetic energy
$\tilde{E}_k=\int_{\tilde{V}}({\textbf{u}}^2/2)d\tilde{V}$
(see figure \ref{fig:mesh1}) and interfacial surface area  $\tilde{A}$ (see figure \ref{fig:mesh2}), respectively, associated with meshes $M_1$ and $M_2$ are essentially indistinguishable highlighting the mesh-independence of our results. 

\begin{figure}
    \centering
    \begin{subfigure}[b]{0.325\textwidth}
    \includegraphics[width=1\linewidth]{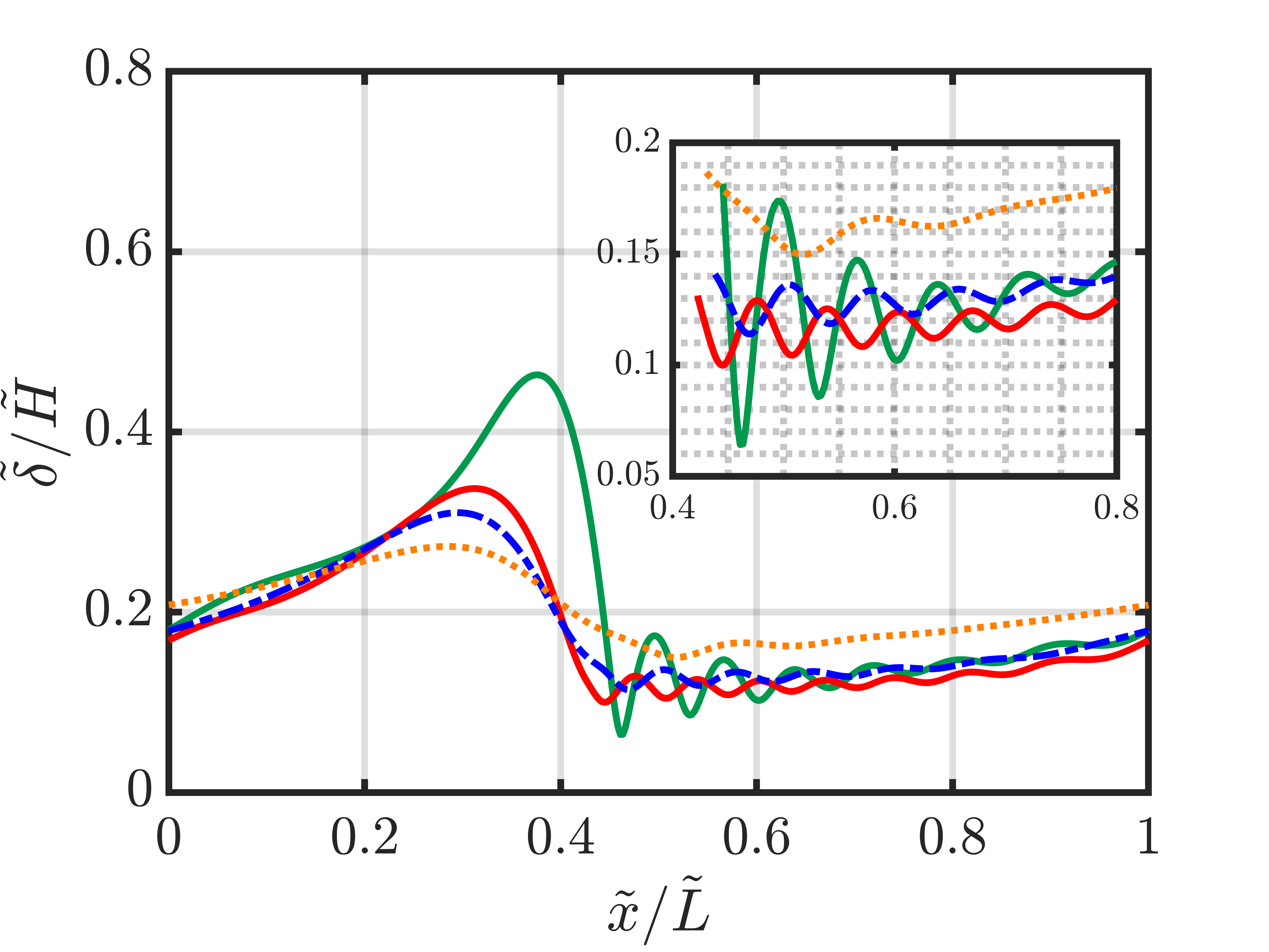}
    \caption{}
    \label{fig:Ma1}
    \end{subfigure}
    \hfill
        \begin{subfigure}[b]{0.325\textwidth}
    \includegraphics[width=1\linewidth]{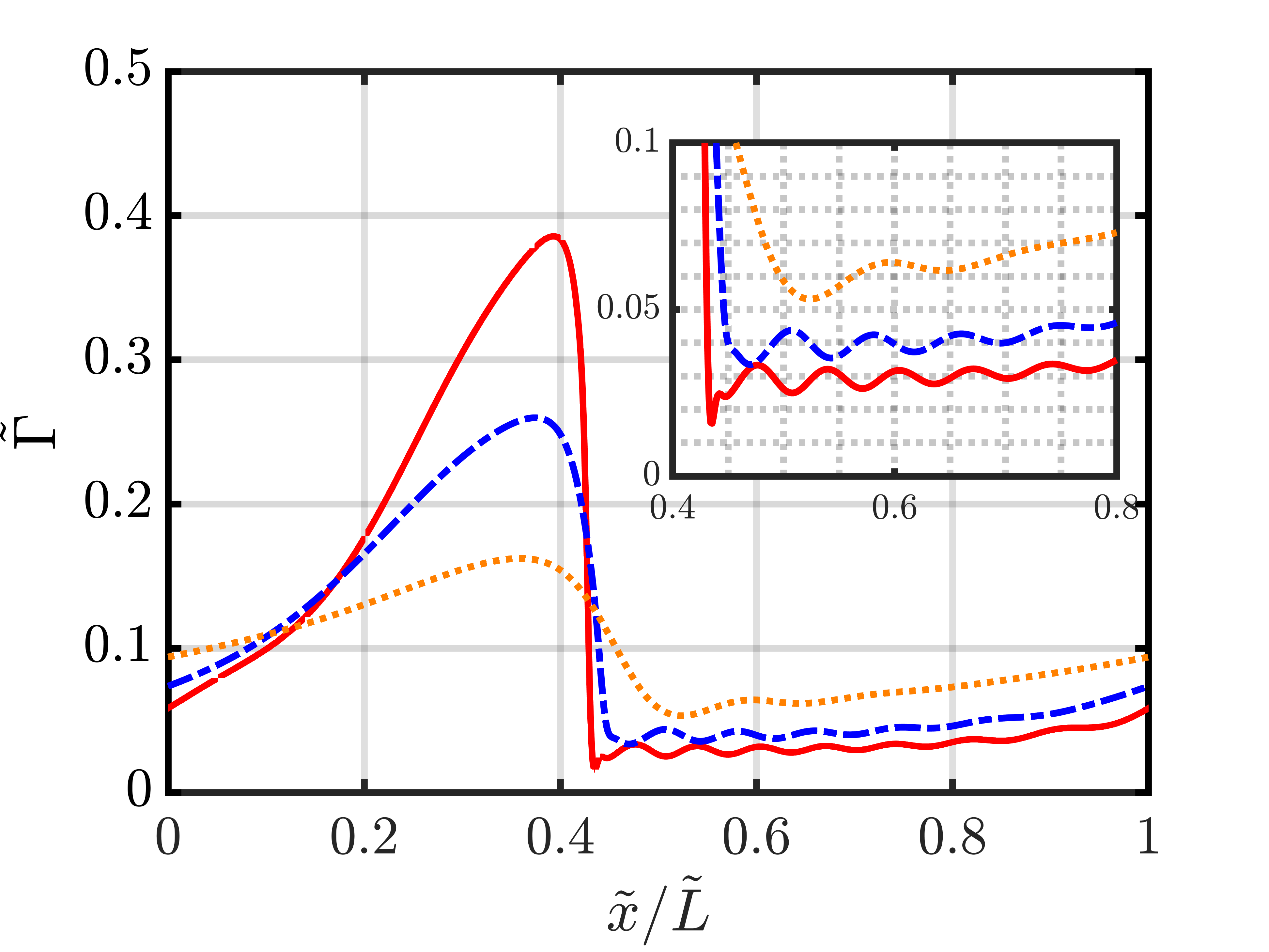}
    \caption{}
    \label{fig:Ma2}
    \end{subfigure}
    \hfill
        \begin{subfigure}[b]{0.325\textwidth}
    \includegraphics[width=1\linewidth]{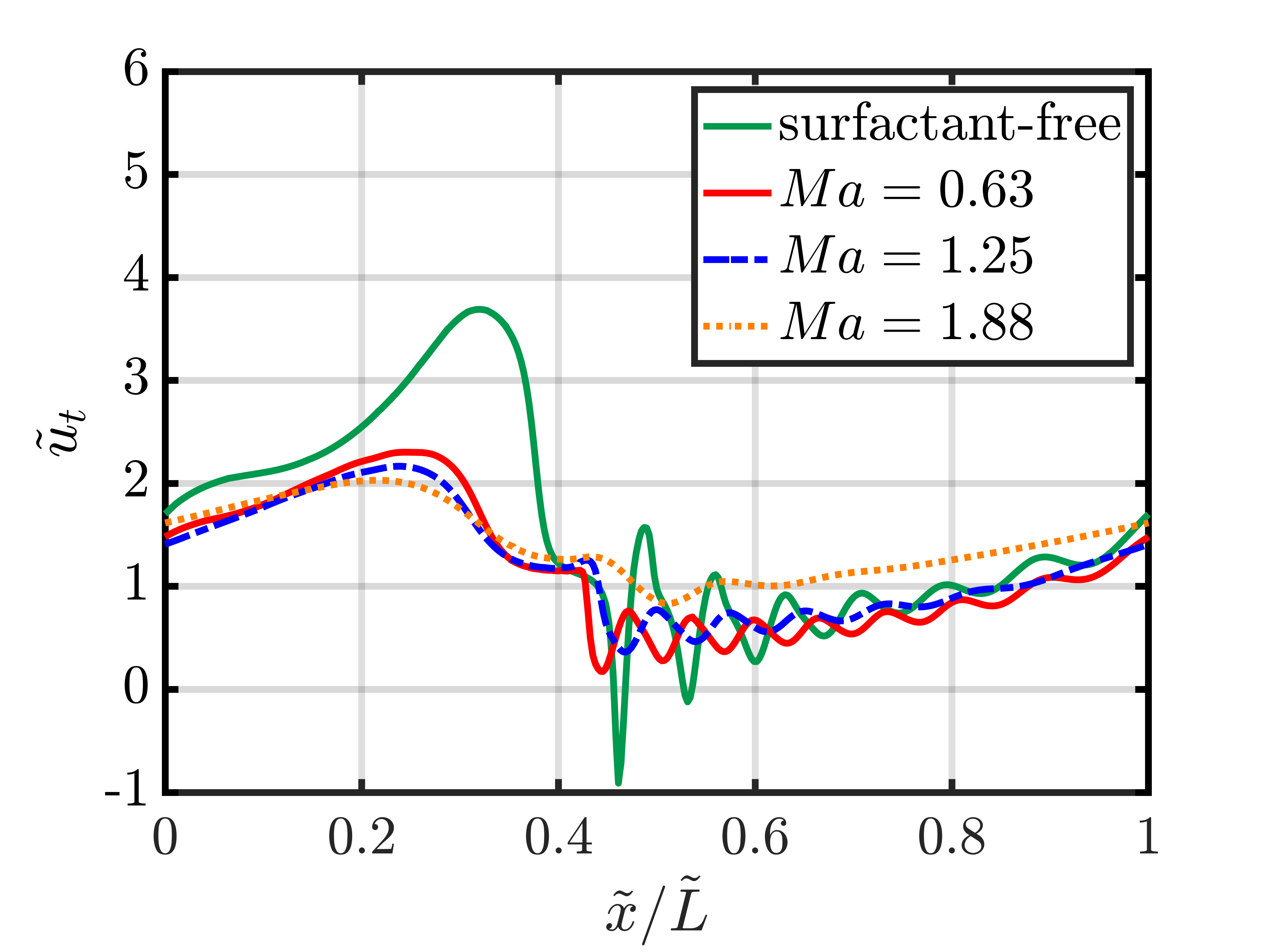}
    \caption{}
    \label{fig:Ma3}
    \end{subfigure}
    \hfill
     \begin{subfigure}[b]{0.49\textwidth}
    \includegraphics[width=0.99\linewidth]{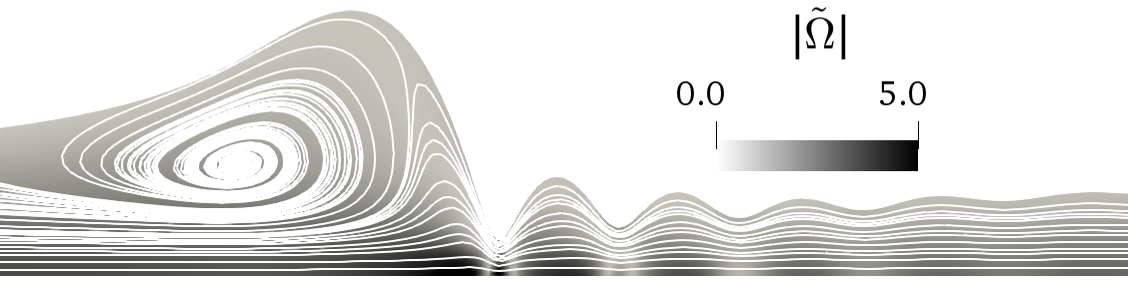}
    \caption{}
    \label{fig:stream1}
    \end{subfigure}
    \hfill
    \begin{subfigure}[b]{0.49\textwidth}
    \includegraphics[width=0.99\linewidth]{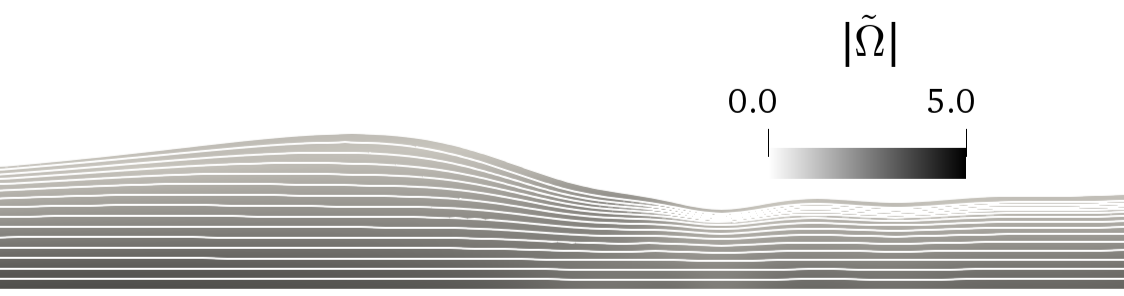}
    \caption{}
    \label{fig:stream2}
    \end{subfigure}
    \hfill
    \caption{Effect of varying the Marangoni parameter on two-dimensional projection in $x-z$ ($y=0$) plane of film thickness, (a),  interfacial surfactant concentration, (b), and the streamwise component of the interfacial velocity, (c), all parameters remaining unchanged from figure \ref{fig2:fig2}. Vortical structure evolution in the reference frame of wave crest for surfactant-free and surfactant-laden films, shown in panels (d) and (e), respectively,  with  the  colour  indicating  the  magnitude of the dimensionless vorticity, $|\tilde\Omega|$. In panel (e), $\Mar=1.88$ and all other parameters remain unchanged from figure \ref{fig2:fig2}.}
    \label{fig:Ma} 
\end{figure}

Following the validation step, we then proceed to add surfactant species to the flow. It should be noted that all surfactant simulations were run until no further topological changes in the interface shape were detected. Although sinusoidal wave segments dominate the first stages of wave development for all studied cases, significant differences in the individual wave evolution stages can be observed in figure \ref{fig2:beta01}-\ref{fig2:beta03} as we increase the Marangoni parameter. For $\Mar=0.63$ (see figure \ref{fig2:beta02}), a horseshoe-shaped wave develops similarly to the surfactant-free case, however, its curvature is smaller and the arc connecting its legs, which bulges upwards for the surfactant-free case, is almost completely flattened. A decrease in the number and increase in wavelength of the capillary wave structures preceding the horseshoe-shaped wave can be seen at $\tilde t=170$ for the surfactant-free (see figure \ref{fig2:clean}) and surfactant-laden (see figure \ref{fig2:beta01}) cases, respectively. The last two panels for $\Mar=0.63$ show a divergence of wave development in comparison to what is observed experimentally for the surfactant-free case. At $\tilde t=625$, we observe that the locations of the horseshoe-shaped and horizontal wave segments are flipped around before the wave stabilises into a quasi two-dimensional shape at $\tilde t=1111$.  An explanation of this wave development, and the underlying role of the surfactants, is provided below. 

\begin{figure}
    \centering
    \begin{subfigure}[b]{0.25\textwidth}
    \includegraphics[width=1\linewidth]{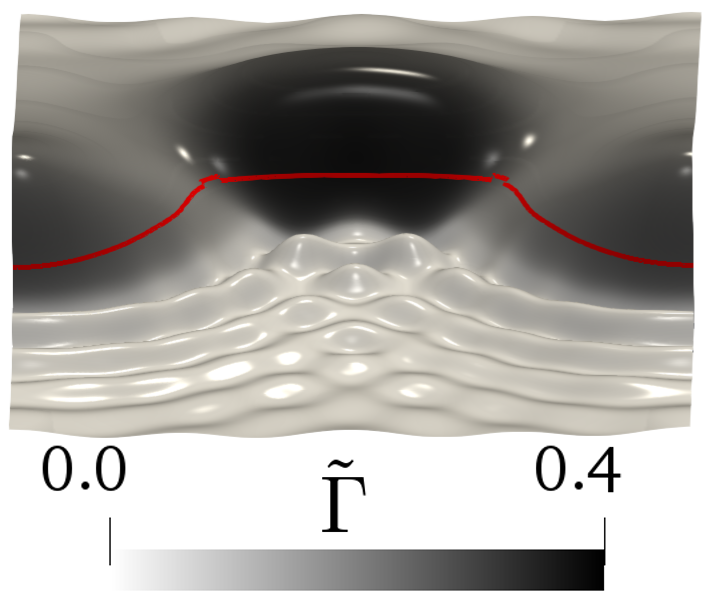}
    \caption{}
    \label{fig:span1}
    \end{subfigure}
    \hfill
        \begin{subfigure}[b]{0.33\textwidth}
    \includegraphics[width=1\linewidth]{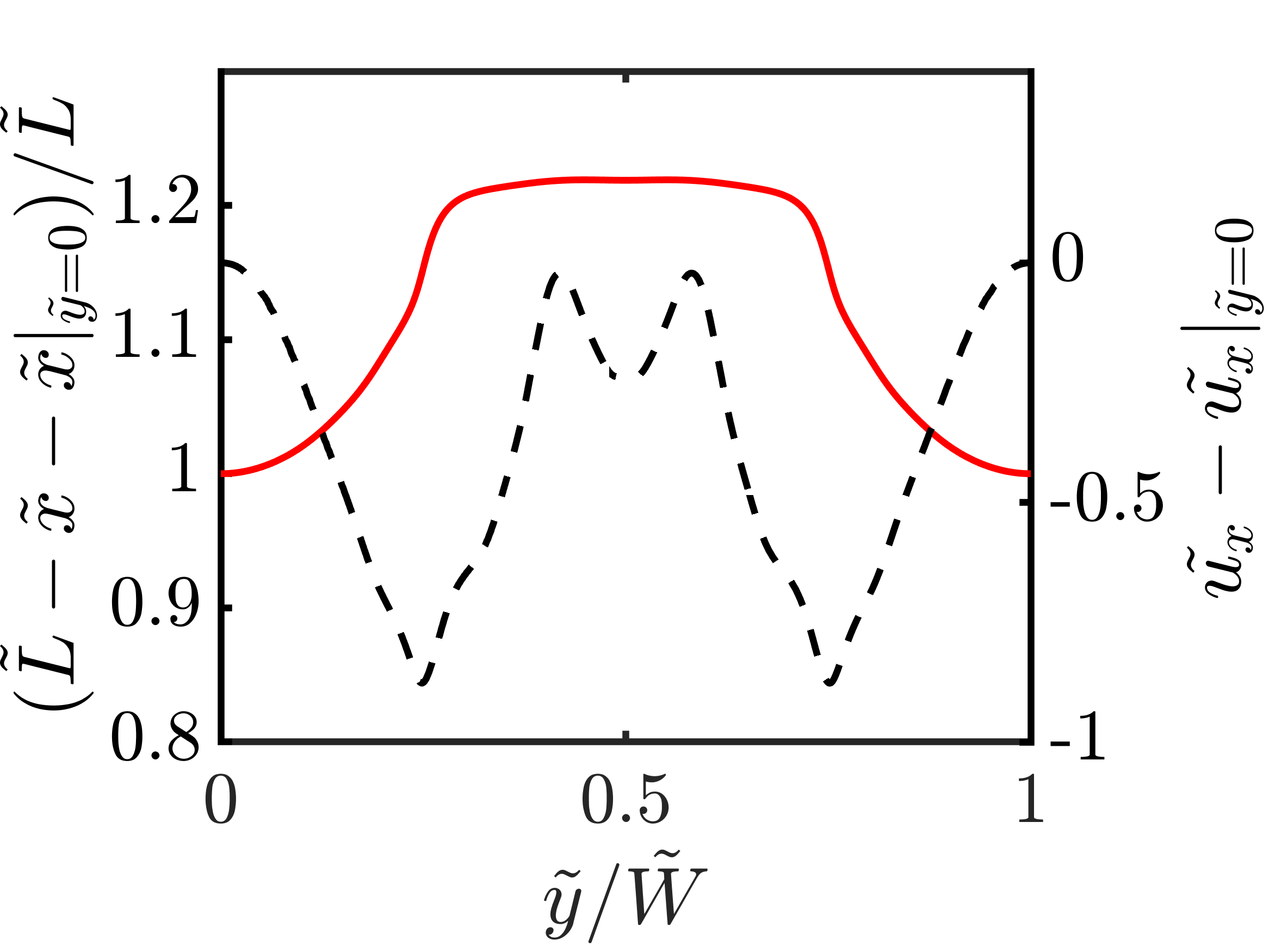}
    \caption{}
    \label{fig:span1ux}
    \end{subfigure}
    \hfill
        \begin{subfigure}[b]{0.33\textwidth}
    \includegraphics[width=1\linewidth]{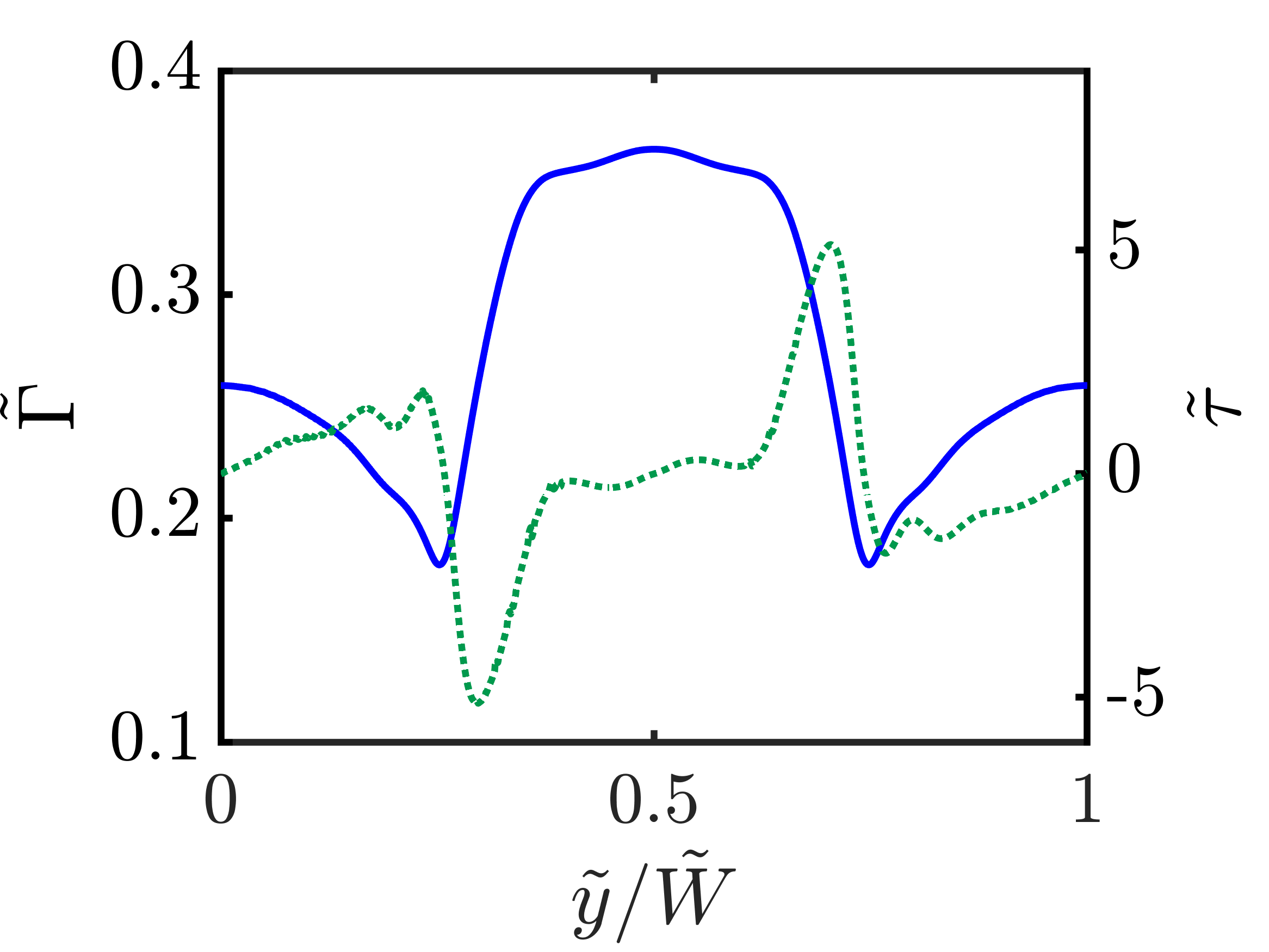}
    \caption{}
    \label{fig:span1tau}
    \end{subfigure}
    \hfill
      \begin{subfigure}[b]{0.25\textwidth}
    \includegraphics[width=1\linewidth]{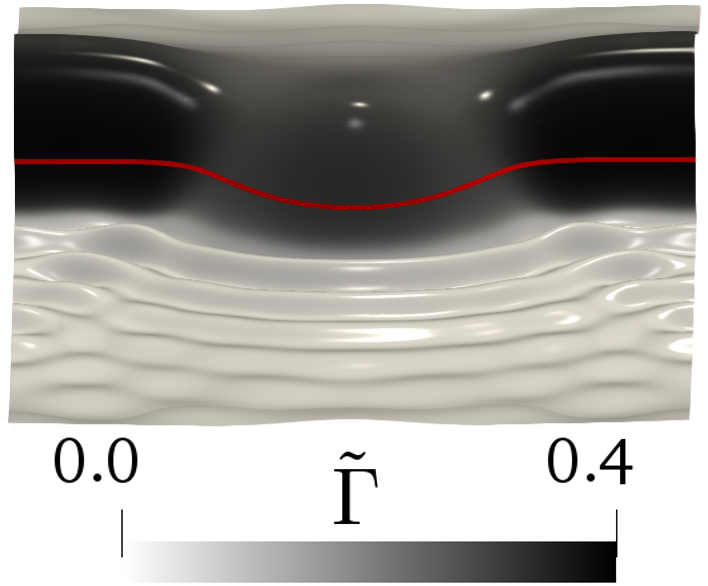}
    \caption{}
    \centering
    \label{fig:span2}
    \end{subfigure}
    \hfill
      \begin{subfigure}[b]{0.33\textwidth}
    \includegraphics[width=1\linewidth]{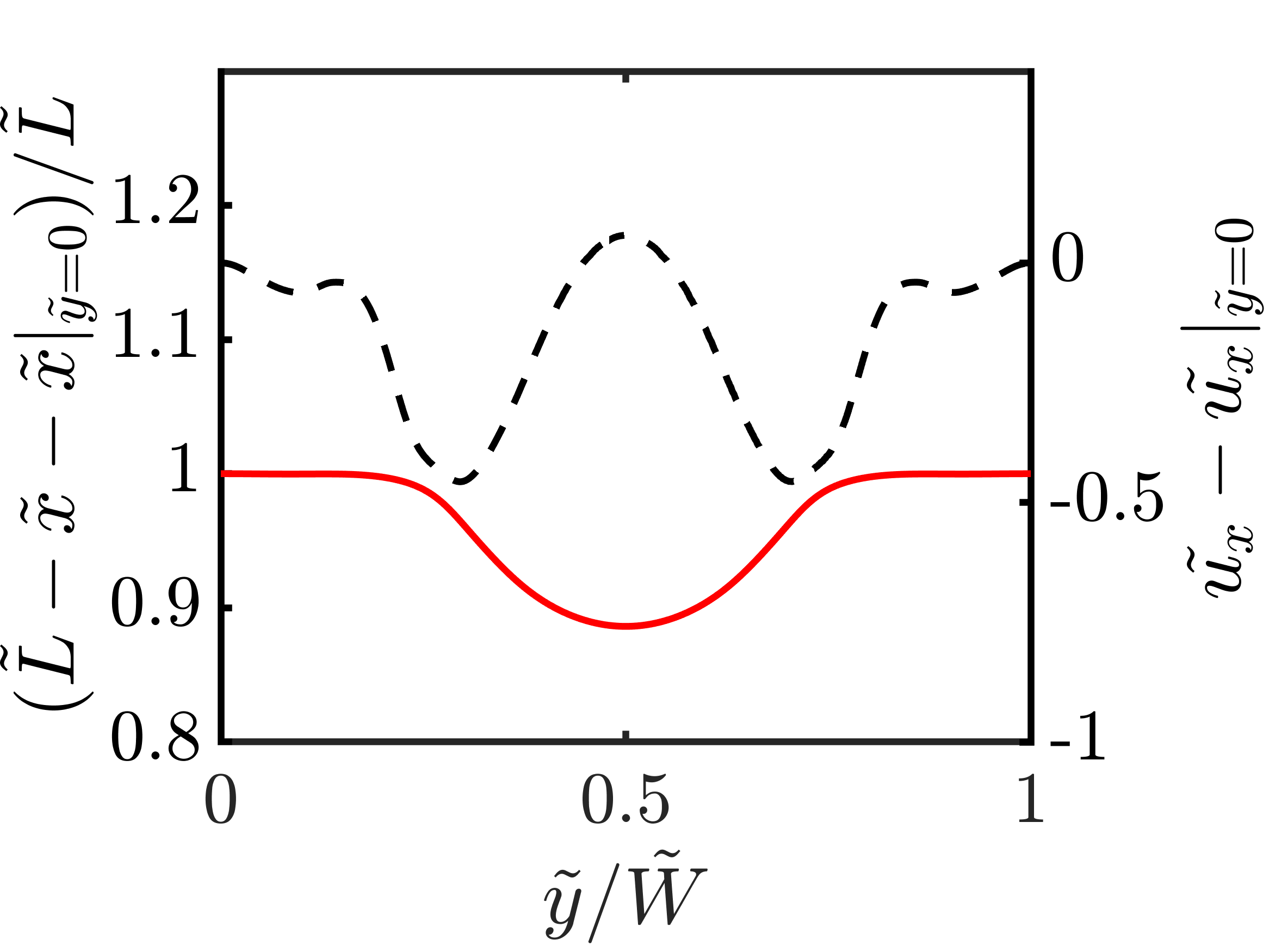}
    \caption{}
    \label{fig:span2ux}
    \centering
    \end{subfigure}
    \hfill
      \begin{subfigure}[b]{0.33\textwidth}
    \includegraphics[width=1\linewidth]{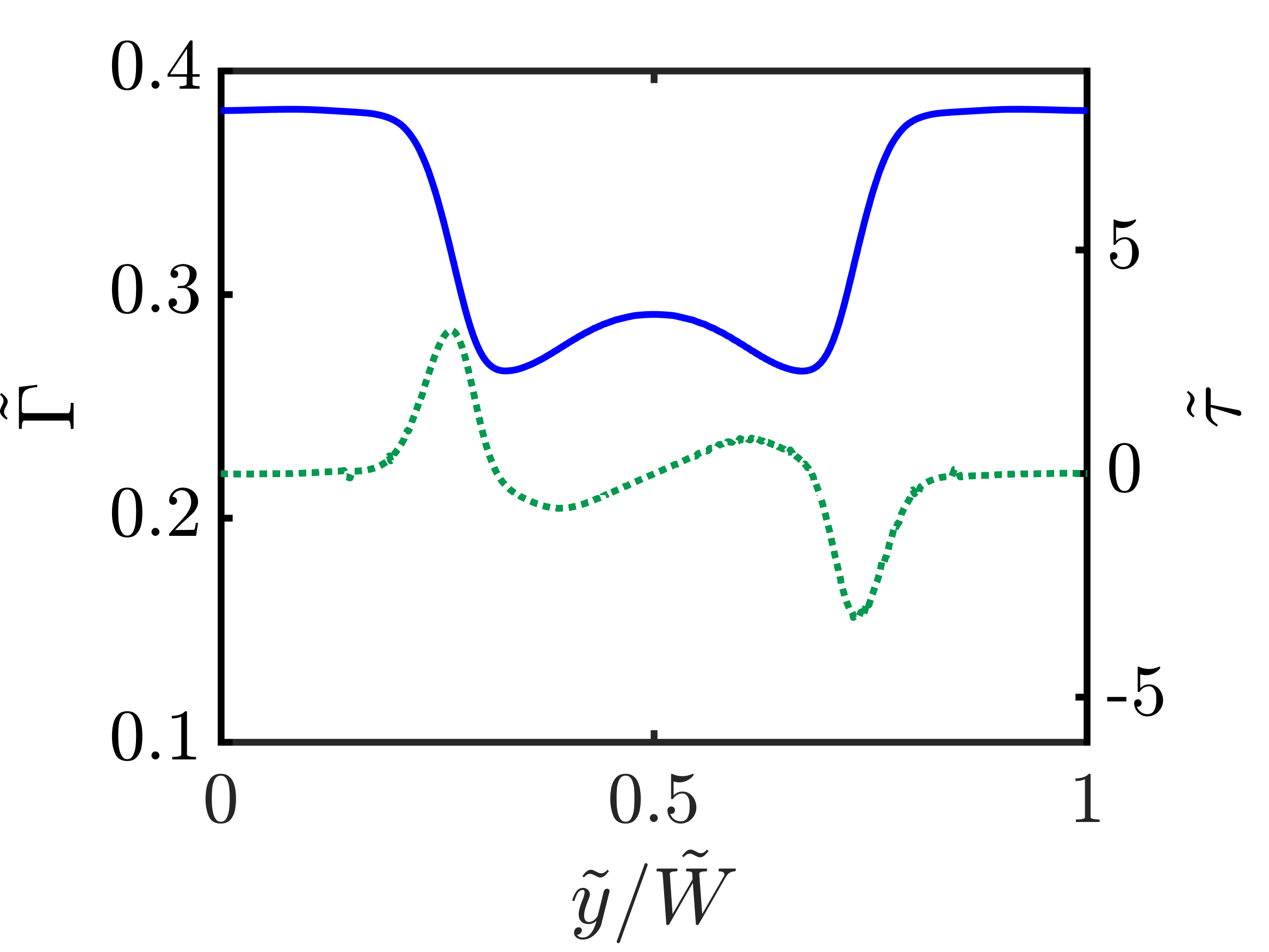}
    \caption{}
    \label{fig:span2tau}
    \centering
    \end{subfigure}
    \hfill
          \begin{subfigure}[b]{0.25\textwidth}
    \includegraphics[width=1\linewidth]{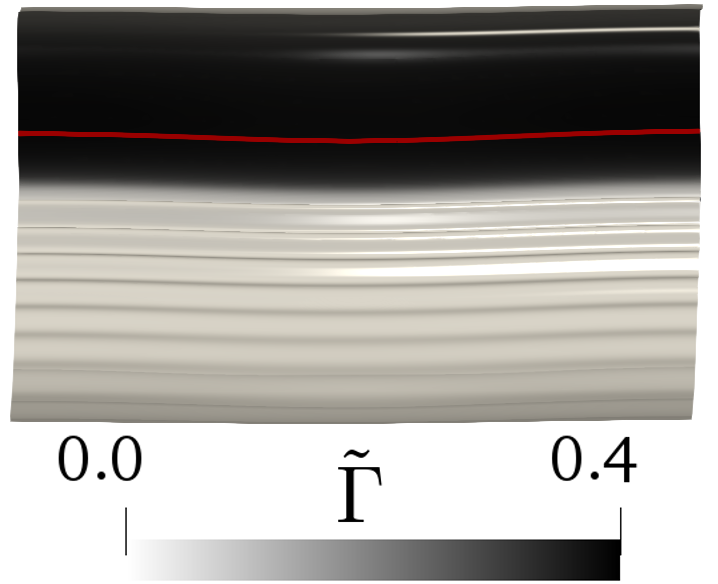}
    \caption{}
    \centering
    \label{fig:span2}
    \end{subfigure}
    \hfill
      \begin{subfigure}[b]{0.33\textwidth}
    \includegraphics[width=1\linewidth]{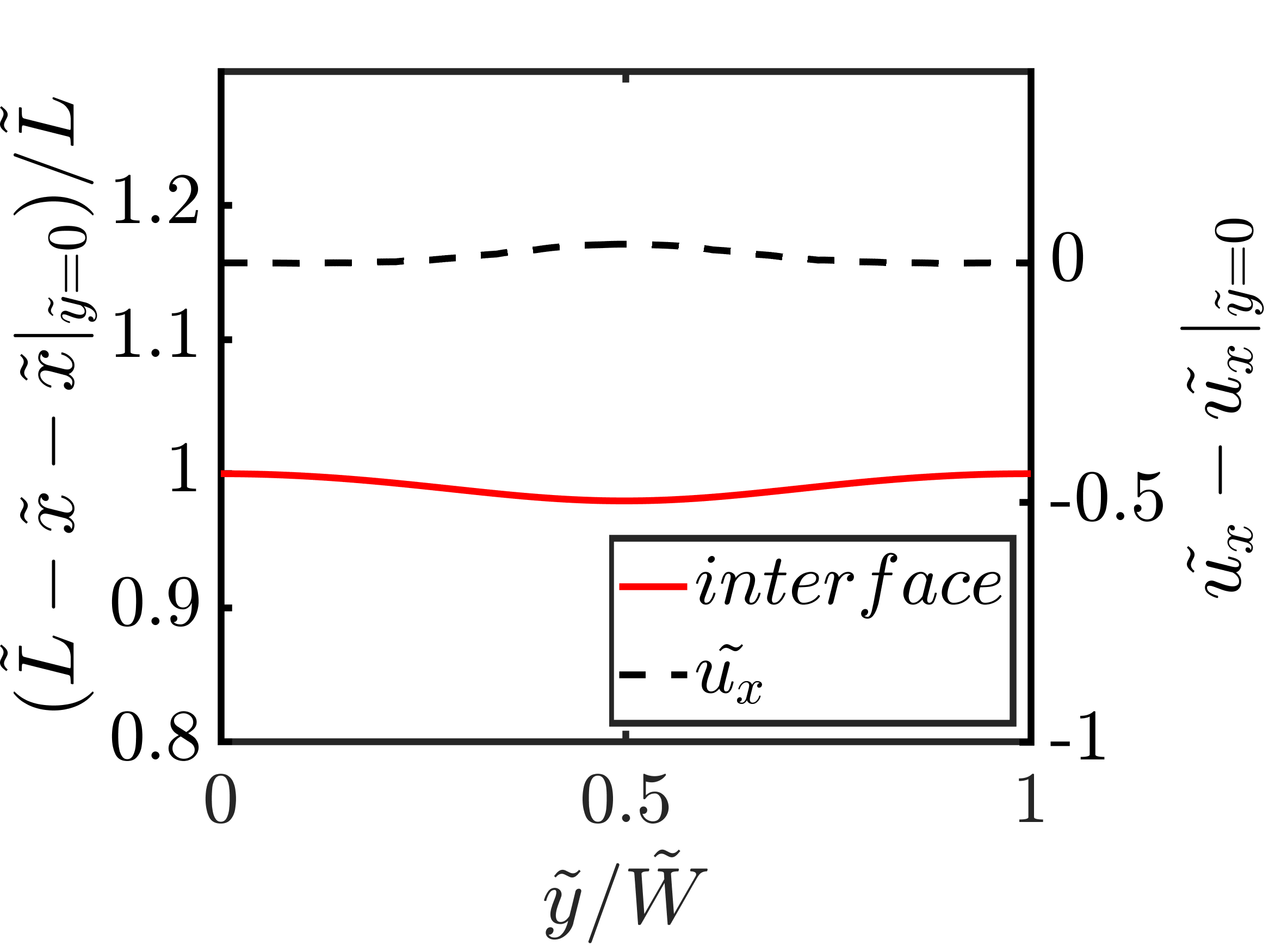}
    \caption{}
    \label{fig:span2ux}
    \centering
    \end{subfigure}
    \hfill
      \begin{subfigure}[b]{0.33\textwidth}
    \includegraphics[width=1\linewidth]{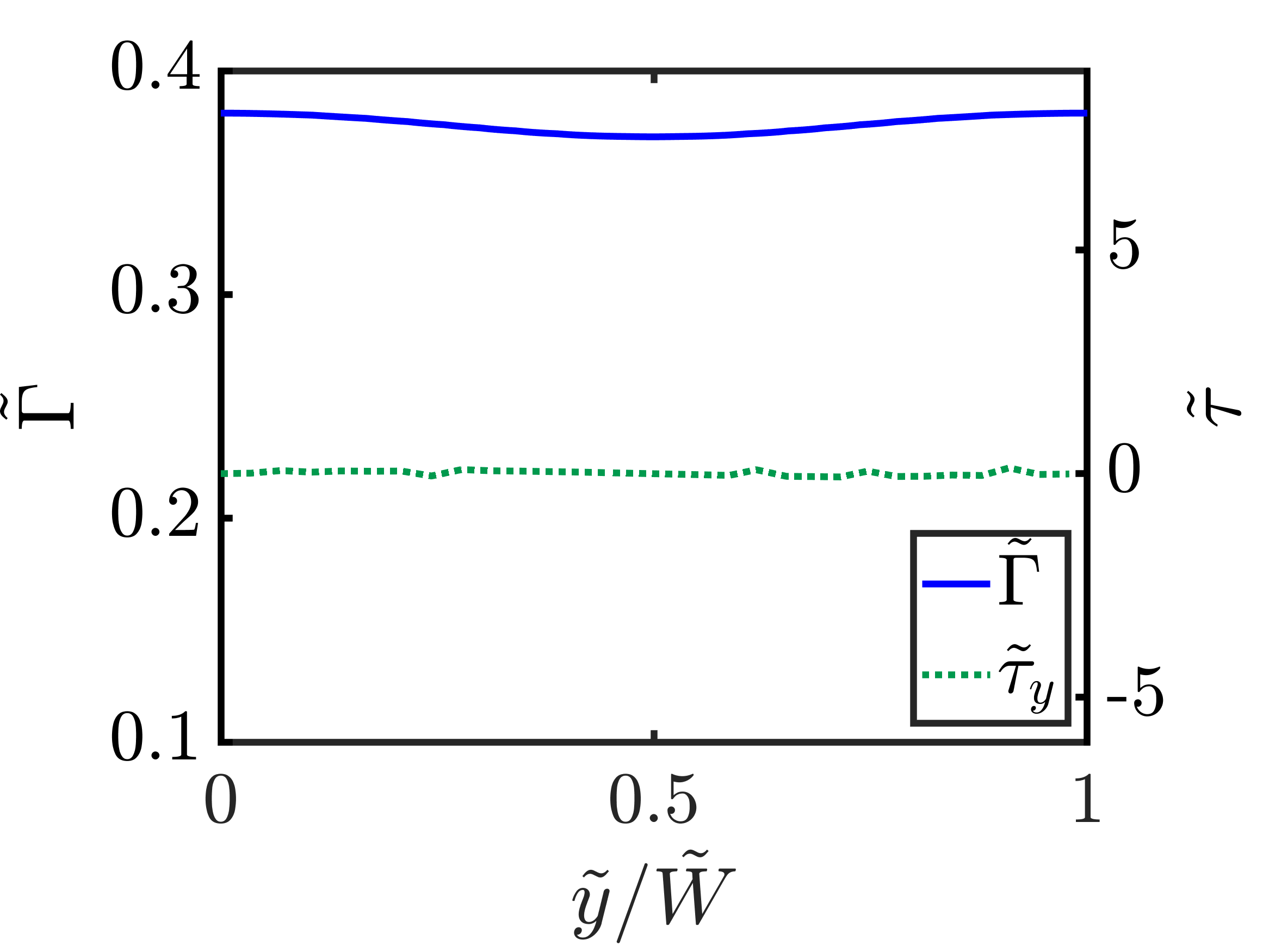}
    \caption{}
    \label{fig:span2tau}
    \centering
    \end{subfigure}
    \hfill
    \caption{3D wave dynamics with the shading indicating the magnitude of surfactant interfacial concentration, $\tilde{\Gamma}$, shown in (a), (d), and (g); 2D projection of interface and streamwise velocity component in the $x-y$ plane ($z=0.2$) shown in (b), (e), and (h); $\tilde{\Gamma}$ and Marangoni stress in $x-y$ plane ($z=0.2$) shown in (c), (f), and (i). Panels (a)-(c), (d)-(f), and (g)-(i) correspond to $\tilde t=223$, $636$, and $1111$, respectively, with $\Mar=0.63$ and the rest of the parameters remain unchanged from figure \ref{fig2:fig2}.}
    \label{fig:span} 
\end{figure}

Attention is now turned towards the $\Mar=1.25$ case presented in figure \ref{fig2:beta02}, where we see that the rise in  
$\Mar$ preserves the initial sinusoidal wave pattern at the early stages of wave development (i.e. up to  $\tilde t=223$) until surfactants act to change the mode of the trailing wave segment from three-dimensional to quasi two-dimensional, by-passing the intermediary step of spanwise oscillation observed for $\Mar=0.63$. For $\Mar=1.25$, we also observe further suppression of capillary wave structures. Finally, we can see the dominant effect of the highest Marangoni parameter on the flow development in figure \ref{fig2:beta03}, where the initially developed sinusoidal shape of the wave is preserved for the entire duration of the simulation, and the capillary wave development is nearly suppressed.

In panels (a)-(c) of figure \ref{fig:Ma}, we examine in greater detail the effect of increasing $Ma$ on the leading capillary structures and trailing wave fronts. The two-dimensional cut performed in the $y=0$ plane reveals that, for all cases investigated, the presence of surfactants suppresses the peak of the trailing wave humps (see figure \ref{fig:Ma1}). This behaviour can be explained further via inspection of the interfacial surfactant concentration in figure \ref{fig:Ma2} where we see that the peak of $\tilde\Gamma$ observed ahead of the wave crest gives rise to a bi-directional Marangoni stress, which drives flow away from the region of high $\tilde\Gamma$ and acts to curb the amplitude of the waves. In figure \ref{fig:Ma3}, the `rigidification' effect induced by the Marangoni stresses is evident by the reduction of $\tilde{u_t}$ of the travelling wave front. We also observe that the suppression of the wave peaks becomes more effective with increasing $\Mar$. 
Upon examination of the vortical structures in panels (d)-(e) of figure \ref{fig:Ma}, we also discover that the presence of surfactant  suppresses actively the re-circulation zone of the trailing hump. Similar effects were observed for all $\Mar$ values examined. 

Next our attention is turned towards the effect of surfactants on the capillary wave structures, where magnification of the region in terms of $\tilde\delta$ and $\tilde\Gamma$ is given in panels \ref{fig:Ma1} and \ref{fig:Ma2}, respectively. For $\Mar=0.63$ and $\Mar=1.25$, the capillary wave structures are still present, however, their amplitude is suppressed significantly. This  Marangoni-driven damping is caused by a local $\tilde\Gamma$ maximum at the peak of each oscillation, which drives the fluid away from the crest. Further evidence in support of the rigidification effect is seen in the decrease of the peak amplitude $\tilde{u_t}$ of each capillary structure (see figure \ref{fig:Ma3}). For $\Mar=1.88$, we observe significant thickening of the capillary region and near complete elimination of the oscillatory structures (viz. the enlarged view inside figure \ref{fig:Ma1}). Further examination of the $\tilde\Gamma$ field in figure \ref{fig:Ma2}, reveals the presence of a concentration gradient that gives rise to a Marangoni stress that drives fluid towards the trough of the trailing hump structure, resulting in the near complete flattening of the overall wave topology. 

We now examine the spanwise oscillatory motion of the wave structures observed for $\Mar=0.63$. In figure \ref{fig:span}, we present snapshots of the three-dimensional wave shape at $\tilde{t}=223$, 636, and 1111 complemented by spanwise two-dimensional representations of the interface, streamwise velocity, $\tilde u_x$ (where the velocity is given in the reference frame of $\tilde u_x$ at $\tilde y=0$), interfacial concentration, $\tilde\Gamma$, and arising spanwise Marangoni stresses,
$\tilde{\tau}$. We observe that the non-uniform distribution of $\tilde\Gamma$ at $\tilde t=223$ (see panels (a)-(c)) gives rise to spanwise Marangoni stresses, which drive fluid flow from the horizontal part of the wave hump towards the legs of the horseshoe-shaped wave and also from the tip of the horseshoe towards its legs. The combined effect of the Marangoni stress is to bridge the gap between the tip of the horseshoe and the horizontal portion of the wave. This effect, however, is sufficiently strong so as to promote the development of the middle portion of the wave segment causing it to accelerate in relation to the adjoining regions giving rise to a spanwise bulge (see panels (d)-(f)). A new local peak in $\tilde\Gamma$, which coincides with the spanwise peak of the bulge, leads to a $\tilde\tau$ structure that induces the final stabilisation of the wave topology (see panels (g)-(i)). Here, the nearly uniform distribution of $\tilde\Gamma$ results in the elimination of all Marangoni stresses.

\begin{figure}
    \centering
    \begin{subfigure}[b]{0.49\textwidth}
    \includegraphics[width=0.99\linewidth]{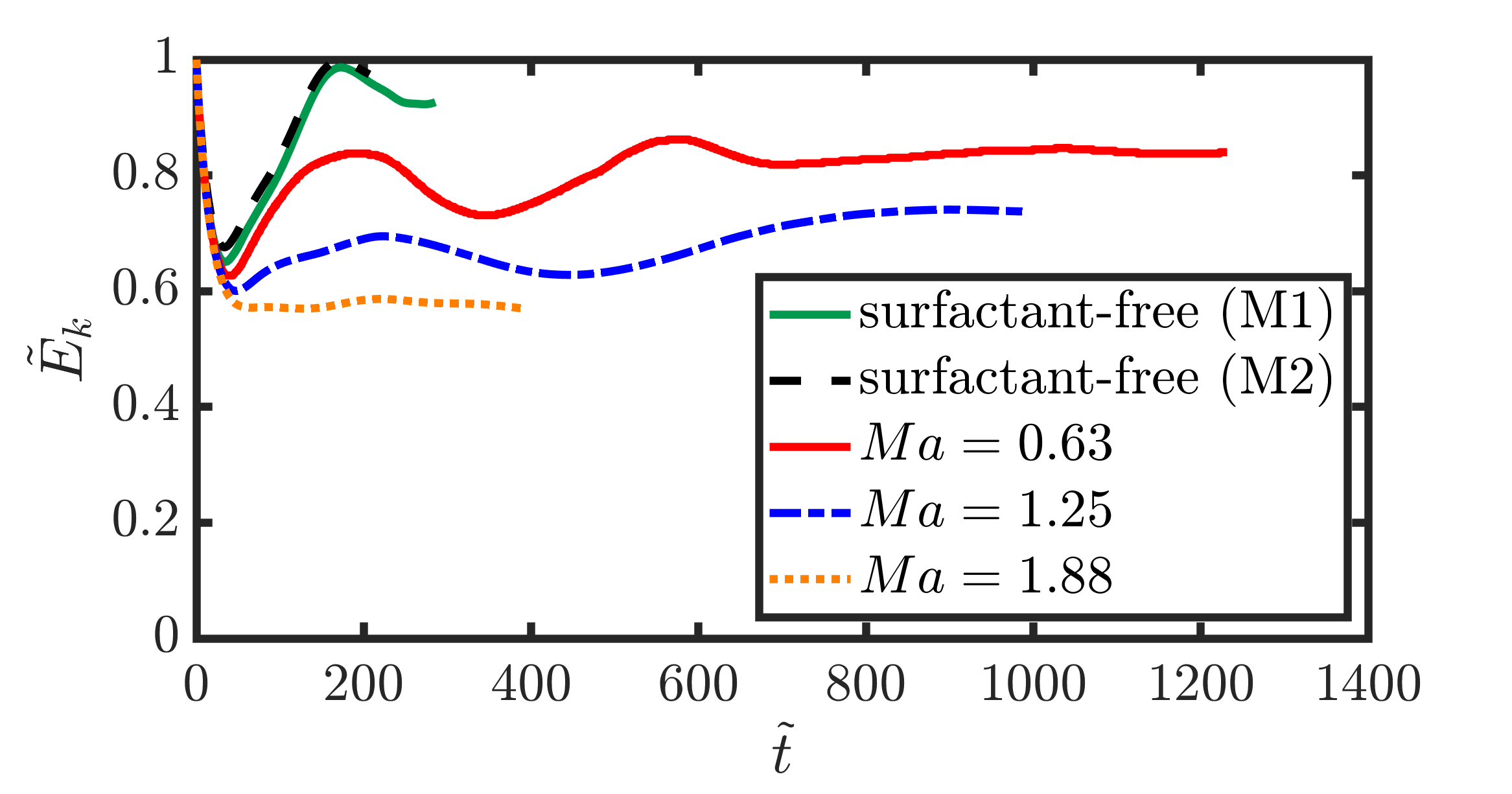}
    \caption{}
    \label{fig:mesh1}
    \end{subfigure}
    \hfill
    \begin{subfigure}[b]{0.49\textwidth}
    \includegraphics[width=0.99\linewidth]{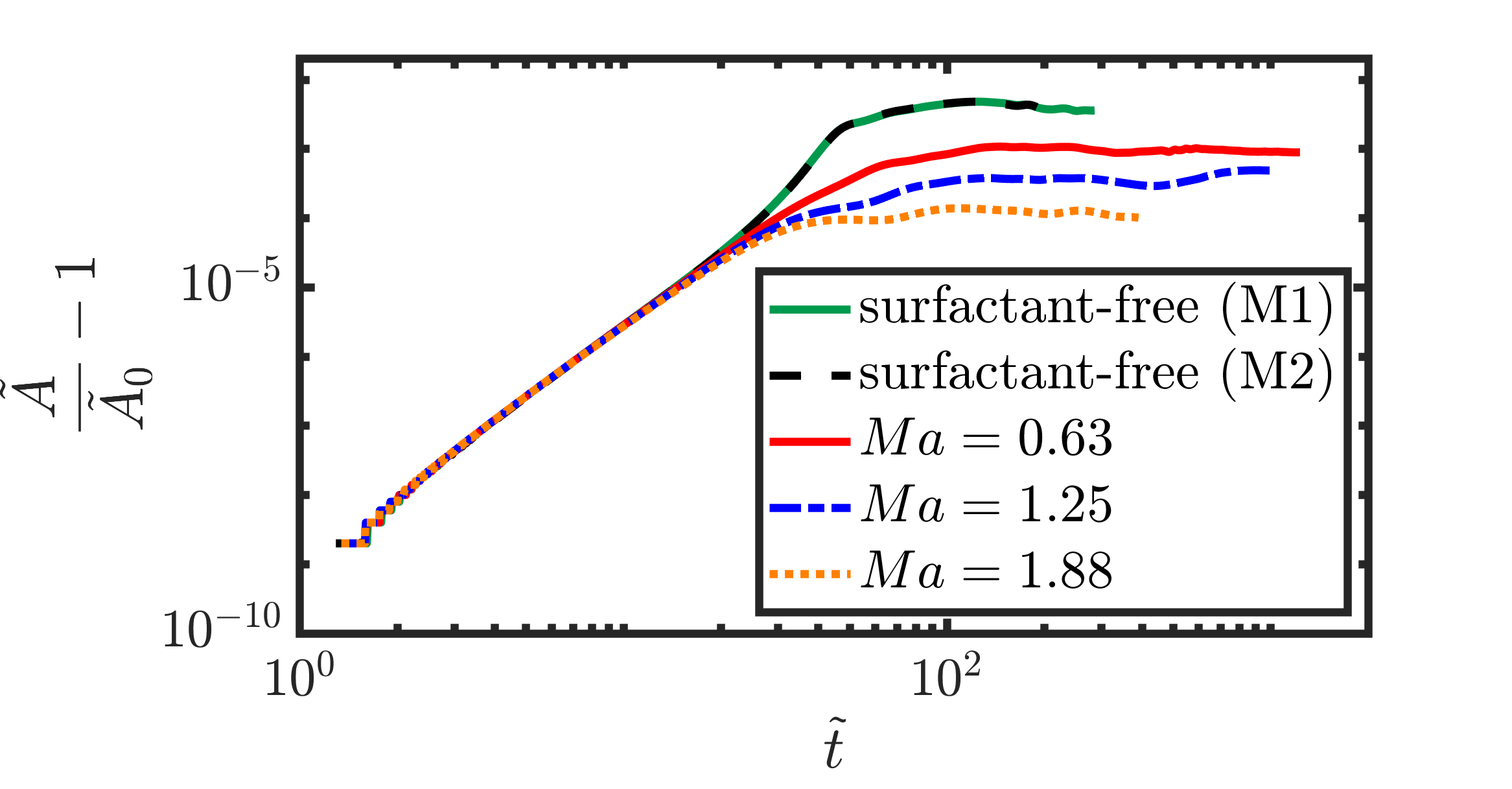}
    \caption{}
    \label{fig:mesh2}
    \end{subfigure}
    \hfill
    \caption{Temporal evolution of (a) kinetic energy, $\tilde E_k$, and (b) surface area, $\tilde A$, scaled on the initial kinetic energy, $E_{k0}$, and  interface area, $A_0$, respectively. The rest of the parameters remain unchanged from figure \ref{fig2:fig2}.}
    \label{fig:mesh} 
\end{figure}

Finally, in figure \ref{fig:mesh}, we show the influence of surfactants on the kinetic energy, defined as $E_k=\int_V (\rho \textbf{u}^2/2) dV$, and the surface area, normalised by their initial values, for the same parameters as in figure \ref{fig2:fig2}. Inspection of the kinetic energy plot in figure \ref{fig:mesh1} reveals that increasing $\Mar$ acts to decrease the overall value of $\tilde E_k$. The amplitude of the  oscillations in $\tilde E_k$ observed at early times for $\Mar=0.63$ is also all but suppressed with increasing $\Mar$. A further increase in $Ma$  to $\Mar=1.88$ rigidifies the flow and eliminates completely any oscillation in $\tilde E_k$. In figure \ref{fig:mesh2}, we see that the presence of surfactant reduces the initial, linear growth rates in interfacial area for all cases, with this effect becoming particularly pronounced at high $\Mar$, in line with the recent observations of  \cite{Hu2020}. 
\section{Concluding remarks}
Three-dimensional numerical simulations of vertically falling liquid films in the presence of insoluble surfactants were carried out for the first time. The numerical predictions for the surfactant-free case were benchmarked against the experimental observations of \cite{Park2003}. For the surfactant-laden case, emphasis was placed on isolating the effect of the Marangoni stresses on the dynamics. The results demonstrate the emergence of oscillations at the wave fronts at low values of the Marangoni parameter, $Ma$, mediated by the Marangoni stresses, brought about by spanwise surfactant concentration gradients; the wave fronts eventually evolve into quasi two-dimensional structures. With increasing $Ma$, the Marangoni stresses led to the progressive elimination of the capillary wave structures  where near complete rigidification, and flattening of the liquid film, were observed for sufficiently large $Ma$. An increase in $Ma$ also resulted in the elimination of vortical structures within the wave crests, and significant reduction in interfacial area, and system kinetic energy. \\

Declaration of Interests. The authors report no conflict of interest. \\


This work is supported by the Engineering $\&$ Physical Sciences Research Council, UK, through the MEMPHIS (EP/K003976/1) and PREMIERE (EP/T000414/1) Programme Grants, and by computing time at HPC facilities provided by the Research Computing Service of Imperial College London (ICL).  The numerical simulations were performed with code BLUE \citep{Shin2017}, and the visualisations were generated using ParaView. The authors also thank Usmaan Farooq (ICL) for fruitful discussions.
 \bibliographystyle{jfm}
 \bibliography{jfm_rapids}

\begin{thebibliography}{33}
\expandafter\ifx\csname natexlab\endcsname\relax\def\natexlab#1{#1}\fi
\def\au#1{#1} \def\ed#1{#1} \def\yr#1{#1}\def\at#1{#1}\def\jt#1{\textit{#1}}
  \def\bt#1{#1}\def\bvol#1{\textbf{#1}} \def\vol#1{#1} \def\pg#1{#1}
  \def\publ#1{#1}\def\arxiv#1{#1}\def\org#1{#1}\def\st#1{\textit{#1}}

\bibitem[Alekseenko {\em et~al.\/}(1994)Alekseenko, Nokariakov, Pokusaev \&
  Fukano]{Alekseenko1994}
{\sc \au{Alekseenko, S.~V.}, \au{Nokariakov, V.~E.}, \au{Pokusaev, B.~G.} \&
  \au{Fukano, T.}} \yr{1994} {\em {Wave flow of liquid films}\/}.  \publ{Begell
  House}.

\bibitem[Benjamin(1964)]{Benjamin1964}
{\sc \au{Benjamin, T.~B.}} \yr{1964}  \at{{Effects of surface contamination on
  wave formation in falling liquid films(Stabilizing effect of surface active
  agents on wave formation in contaminated falling liquid film)}}.
  \jt{Archiwum Mechaniki Stosowanej}  \bvol{16}~(3),  \pg{615--626}.

\bibitem[Bhat \& Samanta(2018)]{Bhat2018}
{\sc \au{Bhat, F.~A.} \& \au{Samanta, A.}} \yr{2018}  \at{{Linear stability of
  a contaminated fluid flow down a slippery inclined plane}}.  \jt{Physical
  Review E}  \bvol{98}~(3),  \pg{1--19}.

\bibitem[Blyth \& Pozrikidis(2004)]{Blyth2004}
{\sc \au{Blyth, M.~G.} \& \au{Pozrikidis, C.}} \yr{2004}  \at{{Effect of
  surfactant on the stability of film flow down an inclined plane}}.
  \jt{Journal of Fluid Mechanics}  \bvol{521},  \pg{241--250}.

\bibitem[Bobylev {\em et~al.\/}(2019)Bobylev, Guzanov, Kvon \&
  Kharlamov]{Bobylev2019}
{\sc \au{Bobylev, A.~V.}, \au{Guzanov, V.~V.}, \au{Kvon, A.~Z.} \&
  \au{Kharlamov, S.~M.}} \yr{2019} {Influence of soluble surfactant on wave
  evolution on falling liquid films}.  \bt{In {\em Journal of Physics:
  Conference Series\/}}, ,  \vol{vol. 1382}.

\bibitem[Chang(1994)]{Chang1994}
{\sc \au{Chang, H.~C.}} \yr{1994}  \at{{Wave Evolution on a Falling Film}}.
  \jt{Annual Review of Fluid Mechanics}  \bvol{26}~(1),  \pg{103--136}.

\bibitem[Chang {\em et~al.\/}(1994)Chang, Cheng, Demekhin \&
  Kopelevich]{Chang1994a}
{\sc \au{Chang, H.~C.}, \au{Cheng, M.}, \au{Demekhin, E.~A.} \& \au{Kopelevich,
  D.~I.}} \yr{1994}  \at{{Secondary and Tertiary Excitation of
  Three-Dimensional Patterns on a Falling Film}}.  \jt{Journal of Fluid
  Mechanics}  \bvol{270},  \pg{251--276}.

\bibitem[Chang {\em et~al.\/}(1996)Chang, Demekhin, Kalaidin \& Ye]{Chang1996}
{\sc \au{Chang, H.~C.}, \au{Demekhin, E.~A.}, \au{Kalaidin, E.} \& \au{Ye, Y.}}
  \yr{1996}  \at{{Coarsening dynamics of falling-film solitary waves}}.
  \jt{Physical Review E - Statistical Physics, Plasmas, Fluids, and Related
  Interdisciplinary Topics}  \bvol{54}~(2),  \pg{1467--1477}.

\bibitem[Cheng \& Chang(1995)]{Cheng1995}
{\sc \au{Cheng, M.} \& \au{Chang, H.~C.}} \yr{1995}  \at{{Competition between
  subharmonic and sideband secondary instabilities on a falling film}}.
  \jt{Physics of Fluids}  \bvol{7}~(1),  \pg{34--54}.

\bibitem[Craster \& Matar(2009)]{Craster2009}
{\sc \au{Craster, R.~V.} \& \au{Matar, O.~K.}} \yr{2009}  \at{{Dynamics and
  stability of thin liquid films}}.  \jt{Reviews of Modern Physics}
  \bvol{81}~(3),  \pg{1131--1198}.

\bibitem[Dietze {\em et~al.\/}(2014)Dietze, Rohlfs, N{\"{a}}hrich, Kneer \&
  Scheid]{Dietze2014}
{\sc \au{Dietze, G.~F.}, \au{Rohlfs, W.}, \au{N{\"{a}}hrich, K.}, \au{Kneer,
  R.} \& \au{Scheid, B.}} \yr{2014}  \at{{Three-dimensional flow structures in
  laminar falling liquid films}}.  \jt{Journal of Fluid Mechanics}  \bvol{743},
   \pg{75--123}.

\bibitem[Georgantaki {\em et~al.\/}(2012)Georgantaki, Vlachogiannis \&
  Bontozoglou]{Georgantaki2012}
{\sc \au{Georgantaki, A.}, \au{Vlachogiannis, M.} \& \au{Bontozoglou, V.}}
  \yr{2012} {The effect of soluble surfactants on liquid film flow}.  \bt{In
  {\em Journal of Physics: Conference Series\/}}, ,  \vol{vol. 395}.

\bibitem[Georgantaki {\em et~al.\/}(2016)Georgantaki, Vlachogiannis \&
  Bontozoglou]{Georgantaki2016}
{\sc \au{Georgantaki, A.}, \au{Vlachogiannis, M.} \& \au{Bontozoglou, V.}}
  \yr{2016}  \at{{Measurements of the stabilisation of liquid film flow by the
  soluble surfactant sodium dodecyl sulfate (SDS)}}.  \jt{International Journal
  of Multiphase Flow}  \bvol{86},  \pg{28--34}.

\bibitem[Hu {\em et~al.\/}(2020)Hu, Fu \& Yang]{Hu2020}
{\sc \au{Hu, T.}, \au{Fu, Q.} \& \au{Yang, L.}} \yr{2020}  \at{{Falling film
  with insoluble surfactants : effects of surface elasticity and surface
  viscosities}}.  \jt{Journal of Fluid Mechanics}  \bvol{889},  \pg{1--19}.

\bibitem[Joo \& Davis(1992)]{Joo1992}
{\sc \au{Joo, S.~W.} \& \au{Davis, S.~H.}} \yr{1992}  \at{{Instabilities of
  three-dimensional viscous falling films}}.  \jt{Journal of Fluid Mechanics}
  \bvol{242}~(529),  \pg{529--547}.

\bibitem[Kalliadasis {\em et~al.\/}(2012)Kalliadasis, Ruyer-Quil, Scheid \&
  Velarde]{Kalliadasis2012}
{\sc \au{Kalliadasis, S.}, \au{Ruyer-Quil, C.}, \au{Scheid, B.} \& \au{Velarde,
  M.~G.}} \yr{2012} {\em Falling Liquid Films\/}.  \publ{Springer}.

\bibitem[Kapitza(1948)]{Kapitza1948}
{\sc \au{Kapitza, P.~L.}} \yr{1948}  \at{{Wave flow of thin layers of viscous
  liquids. II. Flow in a contact with a Gase flux and Heat transfer}}.
  \jt{Zhurnal Eksperimentalnoi i Teoreticheskoi Fiziki}  \bvol{18},
  \pg{3--28}.

\bibitem[Karapetsas \& Bontozoglou(2013)]{Karapetsas2013}
{\sc \au{Karapetsas, G.} \& \au{Bontozoglou, V.}} \yr{2013}  \at{{The primary
  instability of falling films in the presence of soluble surfactants}}.
  \jt{Journal of Fluid Mechanics}  \bvol{729},  \pg{123--150}.

\bibitem[Karapetsas \& Bontozoglou(2014)]{Karapetsas2014}
{\sc \au{Karapetsas, G.} \& \au{Bontozoglou, V.}} \yr{2014}  \at{{The role of
  surfactants on the mechanism of the long-wave instability in liquid film
  flows}}.  \jt{Journal of Fluid Mechanics}  \bvol{741},  \pg{139--155}.

\bibitem[Liu {\em et~al.\/}(1993)Liu, Paul \& Gollub]{Liu1993}
{\sc \au{Liu, J.}, \au{Paul, J.~D.} \& \au{Gollub, J.~P.}} \yr{1993}
  \at{{Measurements of the primary instabilities of film flows}}.  \jt{Journal
  of Fluid Mechanics}  \bvol{250},  \pg{69--101}.

\bibitem[Liu {\em et~al.\/}(1995)Liu, Schneider \& Gollub]{Liu1995}
{\sc \au{Liu, J.}, \au{Schneider, J.~B.} \& \au{Gollub, J.~P.}} \yr{1995}
  \at{{Three-dimensional instabilities of film flows}}.  \jt{Physics of Fluids}
   \bvol{7}~(1),  \pg{55--67}.

\bibitem[Nusselt(1923)]{Nusselt1923}
{\sc \au{Nusselt, W.}} \yr{1923}  \at{{Der W{\"{a}}rmeaustausch am
  Berieselungsk{\"{u}}hler}}.  \jt{Zeitschrift des VDI}  \bvol{67},
  \pg{206--210}.

\bibitem[Oron {\em et~al.\/}(1997)Oron, Davis \& Bankoff]{Oron1997}
{\sc \au{Oron, A.}, \au{Davis, S.~H.} \& \au{Bankoff, S.~G.}} \yr{1997}
  \at{{Long-scale evolution of thin liquid films}}.  \jt{Reviews of Modern
  Physics}  \bvol{69}~(3),  \pg{931--980}.

\bibitem[Park \& Nosoko(2003)]{Park2003}
{\sc \au{Park, C.~D.} \& \au{Nosoko, T.}} \yr{2003}  \at{{Three-Dimensional
  Wave Dynamics on a Falling Film and Associated Mass Transfer}}.  \jt{AIChE
  Journal}  \bvol{49}~(11),  \pg{2715--2727}.

\bibitem[Pereira \& Kalliadasis(2008)]{Pereira2008}
{\sc \au{Pereira, A.} \& \au{Kalliadasis, S.}} \yr{2008}  \at{{Dynamics of a
  falling film with solutal Marangoni effect}}.  \jt{Physical Review E -
  Statistical, Nonlinear, and Soft Matter Physics}  \bvol{78}~(3),  \pg{1--19}.

\bibitem[Scheid {\em et~al.\/}(2006)Scheid, Ruyer-Quil \&
  Manneville]{Scheid2006}
{\sc \au{Scheid, B.}, \au{Ruyer-Quil, C.} \& \au{Manneville, P.}} \yr{2006}
  \at{{Wave patterns in film flows: Modelling and three-dimensional waves}}.
  \jt{Journal of Fluid Mechanics}  \bvol{562},  \pg{183--222}.

\bibitem[Shin {\em et~al.\/}(2017)Shin, Chergui \& Juric]{Shin2017}
{\sc \au{Shin, S.}, \au{Chergui, J.} \& \au{Juric, D.}} \yr{2017}  \at{{A
  solver for massively parallel direct numerical simulation of
  three-dimensional multiphase flows}}.  \jt{Journal of Mechanical Science and
  Technology}  \bvol{31}~(4),  \pg{1739--1751}.

\bibitem[Shin {\em et~al.\/}(2018)Shin, Chergui, Juric, Kahouadji, Matar \&
  Craster]{Shin2018}
{\sc \au{Shin, S.}, \au{Chergui, J.}, \au{Juric, D.}, \au{Kahouadji, L.},
  \au{Matar, O.~K.} \& \au{Craster, R.~V.}} \yr{2018}  \at{{A hybrid interface
  tracking -- level set technique for multiphase flow with soluble
  surfactant}}.  \jt{Journal of Computational Physics}  \bvol{359},
  \pg{409--435}.

\bibitem[Shin \& Juric(2002)]{Shin2002}
{\sc \au{Shin, S.} \& \au{Juric, D.}} \yr{2002}  \at{{Modeling
  Three-Dimensional Multiphase Flow Using a Level Contour Reconstruction Method
  for Front Tracking without Connectivity}}.  \jt{Journal of Computational
  Physics}  \bvol{180}~(2),  \pg{427--470}.

\bibitem[Shin \& Juric(2009)]{Shin_ijnmf_2009}
{\sc \au{Shin, S.} \& \au{Juric, D.}} \yr{2009}  \at{A hybrid interface method
  for three-dimensional multiphase flows based on front tracking and level set
  techniques}.  \jt{Int. J. Num. Meth. Fluids}  \bvol{60},  \pg{753--778}.

\bibitem[Shkadov {\em et~al.\/}(2004)Shkadov, Velarde \& Shkadova]{Shkadov2004}
{\sc \au{Shkadov, V.~Ya.}, \au{Velarde, M.~G.} \& \au{Shkadova, V.~P.}}
  \yr{2004}  \at{{Falling films and the Marangoni effect}}.  \jt{Physical
  Review E - Statistical Physics, Plasmas, Fluids, and Related
  Interdisciplinary Topics}  \bvol{69}~(5),  \pg{15}.

\bibitem[Tailby \& Portalski(1962)]{TailbyS.R.1962}
{\sc \au{Tailby, S.~R.} \& \au{Portalski, S.}} \yr{1962}  \at{{The
  determination of the wavelength on a vertical film of liquid flowing down a
  hydrodynamically smooth plate}}.  \jt{Trans. Inst. Chem. Eng}  \bvol{40},
  \pg{114--122}.

\bibitem[Whitaker(1964)]{Whitaker1964}
{\sc \au{Whitaker, S.}} \yr{1964}  \at{{Effect of surface active agents on the
  stability of falling liquid films}}.  \jt{Industrial and Engineering
  Chemistry Fundamentals}  \bvol{3}~(2),  \pg{132--142}.

\end{thebibliography}

\end{document}